\documentclass[usenatbib]{mn2e}

\usepackage{amsmath,amssymb, graphics}
\usepackage{pgf,tikz}
\usepackage{bm}
\usepackage{graphicx}
\usepackage{epstopdf}
\usepackage{float}
\usepackage{mathrsfs}
\allowdisplaybreaks

\newcommand{\bTs}{{\bm T}_{\rm ds}}
\newcommand{\bTb}{{\bm T}_{\rm db}}
\newcommand{\bT}{{\bm T}}

\newcommand{\oms}{\omega_{\rm ds}}
\newcommand{\omb}{\omega_{\rm db}}
\newcommand{\boms}{\tilde \omega_{\rm ds}}
\newcommand{\bomb}{\tilde \omega_{\rm db}}
\newcommand{\bomd}{\tilde \omega_{\rm sd}}
\newcommand{\omds}{\omega_{\rm ds}}
\newcommand{\omdb}{\omega_{\rm db}}
\newcommand{\bomds}{\tilde \omega_{\rm ds}}
\newcommand{\bomdb}{\tilde \omega_{\rm db}}
\newcommand{\bomsd}{\tilde \omega_{\rm sd}}
\newcommand{\bs}{\bm {\hat s}}
\newcommand{\bl}{\bm {\hat l}}
\newcommand{\ld}{\bm {\hat l}_{\rm d}}
\newcommand{\blone}{{\bm l}_1}
\newcommand{\rin}{r_{\rm in}}
\newcommand{\rout}{r_{\rm out}}
\newcommand{\Sgout}{\Sigma_\text{out}}
\newcommand{\Md}{M_{\rm d}}
\newcommand{\Ld}{L_\text{d}}
\newcommand{\bLd}{{\bm L}_{\rm d}}
\newcommand{\der}{{\rm d}}
\newcommand{\pd}{\partial}
\newcommand{\Om}{\Omega}
\newcommand{\om}{\omega}

\newcommand{\tg}{\theta}
\newcommand{\tgsd}{\theta_{\rm sd}}
\newcommand{\tgsb}{\theta_{\rm sb}}
\newcommand{\tgdb}{\theta_{\rm db}}
\newcommand{\bG}{{\bm G}}
\newcommand{\bText}{{\bm T}_{\rm ext}}
\newcommand{\gb}{g_{\rm b}}
\newcommand{\gs}{g_{\rm s}}
\newcommand{\Vb}{V_{\rm b}}
\newcommand{\Vs}{V_{\rm s}}
\newcommand{\tb}{\tau_{\rm b}}
\newcommand{\ts}{\tau_{\rm s}}
\newcommand{\Wbb}{W_{\rm bb}}
\newcommand{\Wbs}{W_{\rm bs}}
\newcommand{\Wsb}{W_{\rm sb}}
\newcommand{\Wss}{W_{\rm ss}}
\newcommand{\tVb}{\tilde V_{\rm b}}
\newcommand{\tVs}{\tilde V_{\rm s}}
\newcommand{\ttb}{\tilde \tau_{\rm b}}
\newcommand{\tts}{\tilde \tau_{\rm s}}
\newcommand{\tWbb}{\tilde W_{\rm bb}}
\newcommand{\tWbs}{\tilde W_{\rm bs}}
\newcommand{\tWsb}{\tilde W_{\rm sb}}
\newcommand{\tWss}{\tilde W_{\rm ss}}
\newcommand{\cVb}{\mathcal{V}_{\rm b}}
\newcommand{\cVs}{\mathcal{V}_{\rm s}}
\newcommand{\cUb}{\mathcal{U}_{\rm b}}
\newcommand{\cUs}{\mathcal{U}_{\rm s}}
\newcommand{\cWbb}{\mathcal{W}_{\rm bb}}
\newcommand{\cWbs}{\mathcal{W}_{\rm bs}}
\newcommand{\cWsb}{\mathcal{W}_{\rm sb}}
\newcommand{\cWss}{\mathcal{W}_{\rm ss}}
\newcommand{\Vbo}{V_{{\rm b}0}}
\newcommand{\Vso}{V_{{\rm s}0}}
\newcommand{\tbo}{\tau_{{\rm b}0}}
\newcommand{\tso}{\tau_{{\rm s}0}}
\newcommand{\Wbbo}{W_{{\rm bb}0}}
\newcommand{\Wbso}{W_{{\rm bs}0}}
\newcommand{\Wsbo}{W_{{\rm sb}0}}
\newcommand{\Wsso}{W_{{\rm ss}0}}
\newcommand{\cgb}{\gamma_{\rm b}}
\newcommand{\cgs}{\gamma_{\rm s}}
\newcommand{\cgsd}{\gamma_{\rm sd}}
\newcommand{\cgbs}{\gamma_{({\rm bs})} }

\newcommand{\blb}{\bm {\hat l}_{\rm b}}
\newcommand{\hin}{h_{\rm in}}
\newcommand{\hout}{h_{\rm out}}

\newcommand{\bcdot}{{\bm \cdot}}
\newcommand{\btimes}{{\bm \times}}

\newcommand{\ag}{\alpha}
\newcommand{\bg}{\beta}
\newcommand{\cg}{\gamma}
\newcommand{\Sg}{\Sigma}

\newcommand{\Dg}{\Delta}

\newcommand{\brin}{\bar r_\text{in}}
\newcommand{\brout}{\bar r_\text{out}}
\newcommand{\bMd}{\bar M_\text{d}}

\newcommand{\cs}{c_{\rm s}}
\newcommand{\bS}{{\bm S}}
\newcommand{\Mb}{M_{\rm b}}
\newcommand{\bMb}{\bar M_{\rm b}}
\newcommand{\ab}{a_{\rm b}}
\newcommand{\bab}{\bar a_{\rm b}}
\newcommand{\Mst}{M_\star}
\newcommand{\Ms}{M_\star}
\newcommand{\bMst}{\bar M_\star}
\newcommand{\Rst}{R_\star}
\newcommand{\Rs}{R_\star}
\newcommand{\bRst}{\bar R_\star}
\newcommand{\Omst}{\Omega_\star}
\newcommand{\bOmst}{\bar \Omega_\star}
\newcommand{\kq}{k_{\rm q}}

\newcommand{\ks}{k_\star}
\newcommand{\Msun}{{\rm M}_\odot}
\newcommand{\Rsun}{{\rm R}_\odot}

\newcommand{\be}{\begin{equation}}
\newcommand{\ee}{\end{equation}}

\begin{document}
\title[Disk Warp in Star-Disk-Binary Systems]
{Effects of Disk Warping on the Inclination Evolution of Star-Disk-Binary Systems}

\author[J. J. Zanazzi and Dong Lai]{J. J. Zanazzi$^{1}$\thanks{Email: jjz54@cornell.edu}, and Dong Lai$^{1}$ \\
$^{1}$Cornell Center for Astrophysics and Planetary Science, Department of Astronomy, Cornell University, Ithaca, NY 14853, USA}

\maketitle
\begin{abstract}
Several recent studies have suggested that circumstellar disks in
young stellar binaries may be driven into misalignement with their
host stars due to secular gravitational interactions between the star, disk
and the binary companion. The disk in such systems is twisted/warped
due to the gravitational torques from the oblate central star and the
external companion.  We calculate the disk warp profile, taking into
account of bending wave propagation and viscosity in the disk. We show that for
typical protostellar disk parameters, the disk warp is small, thereby 
justifying the ``flat-disk'' approximation adopted in previous
theoretical studies. However, the viscous dissipation associated with
the small disk warp/twist tends to drive the disk toward alignment with the
binary or the central star.  We calculate the relevant timescales for the alignment. 
We find the alignment is effective for sufficiently cold disks with strong external torques, especially for systems with rapidly rotating stars,
but is ineffective for the majority of star-disk-binary systems.  Viscous warp driven alignment may be necessary to account for the observed spin-orbit alignment in multi-planet systems
if these systems are accompanied by an inclined binary companion.
\end{abstract}


\section{Introduction}
\label{sec:Intro}

Circumstellar disks in young protostellar binary systems are likely
to form with an inclined orientation relative to the binary orbital
plane, as a result of the complex star/binary/disc formation processes
(e.g.  \citealt{Bate(2003),McKeeOstriker(2007),Klessen(2011)}).
Indeed, many misaligned circumstellar disks in protostellar binaries have been
found in recent years (e.g. \citealt{Stapelfeldt(1998),Stapelfeldt(2003),Neuhauser(2009),JensenAkeson(2014),Williams(2014),Brinch(2016),Fernandez-Lopez(2017),Lee(2017)}).  Such misaligned disks experience
differential gravitational torques from the binary companion, and are
expected to be twisted/warped while undergoing damped precession
around the binary (e.g. \citealt{LubowOgilvie(2000),Bate(2000),FoucartLai(2014)}).
On the other hand, a spinning protostar has a rotation-induced quadrupole, 
and thus exerts a torque on the disk (and also receives a back-reaction torque)
when the stellar spin axis and the disk axis are misaligned. This torque
tends to induce warping in the inner disk and drives mutual precession
between the stellar spin and disk.
In the presence of both torques on the disk, from the binary and from the central star,
how does the disk warp and precess? What is the long-term evolution of 
the disk and stellar spin in such star-disk-binary systems? These are the questions 
we intend to address in this paper.

Several recent studies have examined the secular dynamics of the
stellar spin and circumstellar disk in the presence of an inclined
binary companion \citep{Batygin(2012),BatyginAdams(2013),Lai(2014),SpaldingBatygin(2014),SpaldingBatygin(2015)}. These studies were motivated by the observations of
spin-orbit misalignments in exoplanetary systems containing hot
Jupiters, i.e., the planet's orbital plane is often misaligned with
the stellar rotational equator (see \citealt{WinnFabrycky(2015)} and
\citealt{Triaud(2017)} for recent reviews).  It was shown that
significant ``primordial'' misalignments may be generated while the
planetary systems are still forming in their natal protoplanetary
disks through secular star-disk-binary gravitational
interactions \citep{BatyginAdams(2013),Lai(2014),SpaldingBatygin(2014),SpaldingBatygin(2015)}.
In these studies, various assumptions were made about the star-disk
interactions, and uncertain physical processes such as star/disk winds,
magnetic star-disk interactions, and accretion of disk angular
momentum onto the star were incorporated in a parameterized manner.
Nevertheless, the production of spin-orbit misalignments seems quite 
robust.

In \cite{ZanazziLai(2017b)}, we showed that the formation of hot
Jupiters in the protoplanetary disks can significantly suppress the
excitation of spin-orbit misalignment in star-disk-binary
systems. This is because the presence of such close-in giant planets
lead to strong spin-orbit coupling between the planet and its host
star, so that the spin-orbit misalignment angle is adiabatically
maintained despite the gravitational perturbation from the binary
companion. However, the formation of small planets or
distant planets (e.g. warm Jupiters) do not affect the generation of primordial
misalignments between the host star and the disk.

A key assumption made in all previous studies on misalignments in
star-disk-binary systems \citep{BatyginAdams(2013),Lai(2014),SpaldingBatygin(2015)} is
that the disk is nearly flat and behaves like a rigid plate in
response to the external torques from the binary and from the host
star.  The rationale for this assumption is that different regions of
the disk can efficiently communicate with each other through
hydrodynamical forces and/or self-gravity, such that the disk
stays nearly flat. However, to what extent this assumption is valid is
uncertain, especially because in the star-disk-binary system
the disk experiences two distinct torques from the
oblate star and from the binary which tend to drive the disk toward
different orientations (see \citealt{TremaineDavis(2014)} for examples of
non-trivial disk warps when a disk is torqued by different
forces). Moreover, the combined effects of disk warps/twists (even if
small) and viscosity can lead to non-trivial long-term evolution of the
star-disk-binary system. Previous works on warped disks in the bending wave regime have considered a single external torque, such as an ext binary companion \citep{LubowOgilvie(2000),Bate(2000),FoucartLai(2014)}, an inner binary \citep{Facchini(2013),LodatoFacchini(2013),FoucartLai(2014),ZanazziLai(2018)}, magnetic torques from the central star \citep{FoucartLai(2011)}, a central spinning black hole \citep{DemianskiIvanov(1997),Lubow(2002),Franchini(2016),CharkrabortyBhattacharyaa(2017)}, and a system of multiple planets on nearly coplanar orbits \citep{LubowOgilvie(2001)}.
In this paper, we will focus on the
hydrodynamics of warped disks in star-disk-binary systems, and will
present analytical calculations for the warp amplitudes/profiles and
the rate of evolution of disk inclinations due to viscous dissipation
associated with these warps/twists.


This paper is organized as follows. Section~\ref{sec:Setup} describes
the setup and parameters of the star-disk-binary system we study.
Section~\ref{sec:DiskWarp} presents all the technical calculations of
our paper, including the disk warp/twist profile and effect of viscous
dissipation on the evolution of system.  Section~\ref{sec:Dyn}
examines how viscous dissipation from disk warps modifies the
long-term evolution of star-disk-binary systems.
Section~\ref{sec:Discuss} discusses theoretical uncertainties of our
work.  Section~\ref{sec:Conc} contains our conclusions.

\section{Star-Disk-Binary System and Gravitational Torques}
\label{sec:Setup}

Consider a central star of mass $\Mst$, radius $\Rst$, rotation rate $\Omst$, with a circumstellar disk of mass $\Md$, and inner and outer truncation radii of $\rin$ and $\rout$, respectively.  This star-disk system is in orbit with a distant binary companion of mass $\Mb$ and semimajor axis $\ab$.  The binary companion exerts a torque on the disk, driving it into differential precession around the binary angular momentum axis $\blb$.  Averaging over the orbital period of the disk annulus and binary, the torque per unit mass is
\be
\bT_{\rm db} = -r^2 \Om \omb (\bl \bcdot \blb) \blb \btimes \bl,
\label{eq:Tdb}
\ee
where $\Om(r) \simeq \sqrt{G \Mst/r^3}$ is the disk angular frequency, $\bl = \bl(r,t)$ is the unit orbital angular momentum axis of a disk ``ring" at radius $r$, and
\be
\om_{\rm db}(r) = \frac{3 G \Mst}{4 \ab^3 \Om}
\label{eq:omdb}
\ee
is the characteristic precession frequency of the disk ``ring'' at radius $r$.  Similarly, the rotation-induced stellar quadrapole drives the stellar spin axis $\bs$ and the disk onto mutual precession.  The stellar rotation leads to a difference in the principal components of the star's moment of inertia of $I_3 - I_1 = \kq \Mst \Rst^2 \bOmst^2$, where $\kq \simeq 0.1$ for fully convective stars \citep{Lai(1993)}.  Averaging over the orbital period of the disk annulus, the torque on the disk from the oblate star is
\be
\bTs(r,t) = -r^2 \Om \oms (\bs \bcdot \bl)\bs \btimes \bl,
\label{eq:Tds}
\ee
where
\be
\omds(r) = \frac{3 G (I_3 - I_1)}{2 r^5 \Om} = \frac{3 G \kq \Mst \Rst^2 \bOmst^2}{2 r^5 \Om}
\label{eq:omds}
\ee
is the characteristic precession frequency of the disk ring at radius $r$.  Since $\omdb$ and $\omds$ both depend on $r$, the disk would quickly lose coherence if there were no internal coupling between the different ``rings.''

We introduce the following rescaled parameters typical of protostellar systems:
\begin{align}
&\bMst = \frac{\Mst}{1 \, \Msun},
\hspace{3mm}
\bRst = \frac{\Rst}{2 \, \Rsun},
\hspace{3mm}
\bOmst = \frac{\Omst}{\sqrt{G \Mst/\Rst^3}},
\nonumber \\
&\bMd = \frac{\Md}{0.1 \, \Msun},
\hspace{3mm}
\brin = \frac{\rin}{8 \, \Rsun},
\hspace{3mm}
\brout = \frac{\rout}{50 \, \text{au}},
\nonumber \\
&\bMb = \frac{\Mb}{1 \, \Msun},
\hspace{5mm}
\bab = \frac{\ab}{300 \, \text{au}}.
\label{eq:pars}
\end{align}
 The rotation periods of T Tauri stars vary from $P_\star \sim 1 - 10 \ \text{days}$ \citep{Bouvier(2013)}, corresponding to $\bOmst \sim 0.3 - 0.03$.  We fix the canonical value of $\bOmst$ to be $0.1$, corresponding to a stellar rotation period of $P_\star = 3.3 \, \text{days}$.  The other canonical values in Eq.~\eqref{eq:pars} are unity, except the disk mass, which can change significantly during the disk lifetime.  
 Our choice of $\rin$ is motivated by typical values of a T Tauri star's magnetospheric radius $r_{\rm m}$, set by the balance of magnetic and plasma stresses (see \citealt{Lai(2014b)} for review)
 \begin{align}
 &\rin \approx r_{\rm m} = \eta \left( \frac{\mu_\star^4}{G \Ms \dot M^2} \right)^{1/7}
 \nonumber \\
 &= 7.4 \, \eta \left( \frac{B_\star}{1 \, \text{kG}} \right)^{4/7} \left( \frac{10^{-7} \, \Msun/\text{yr}}{\dot M} \right)^{2/7} \frac{\bRst^{12/7}}{\bMst^{1/7}} \Rsun.
 \label{eq:rm}
 \end{align}
Here, $\mu_\star = B_\star \Rst^3$ is the stellar dipole moment, $B_\star$ is the stellar magnetic field, $\dot M$ is the accretion rate onto the central star (e.g. \citealt{Rafikov(2017)}), and $\eta$ is a parameter of order unity.
We note that we take the stellar radius to be fixed, in contrast to the models of \cite{BatyginAdams(2013)} and \cite{SpaldingBatygin(2014),SpaldingBatygin(2015)}, but we argue this will not change our results significantly.  We are primarily concerned with the effects of viscous dissipation from disk warping, and a changing stellar radius will not affect the viscous torque calculations to follow.

We parameterize the disk surface density $\Sg = \Sg(r,t)$ as
\be
\Sg(r,t) = \Sgout(t) \left( \frac{\rout}{r} \right)^p.
\label{eq:p} 
\ee
We take $p=1$ unless otherwise noted.  The disk mass $\Md$ is then (assuming $\rin \ll \rout$)
\be
\Md = \int_{\rin}^{\rout} 2 \pi \Sg r \der r \simeq \frac{2\pi \Sgout \rout^2}{2-p} .
\label{eq:Md}
\ee
The disk angular momentum vector is $\bLd = \Ld \ld$ (assuming a small disk warp), and stellar spin angular momentum vector is $\bS = S \bs$, where $\ld$ and $\bs$ are unit vectors, and
\begin{align}
\Ld &= \int_{\rin}^{\rout} 2 \pi \Sg r^3 \Om \der r \simeq \frac{2-p}{5/2-p} \Md  \sqrt{G \Mst \rout}, \\
S &= k_\star \Mst \Rst^2 \Omst.
\end{align}
Here $k_\star \simeq 0.2$ for fully convective stars ( e.g. \citealt{Chandrasekhar(1939)}).  The binary has orbital angular momentum ${\bm L}_{\rm b} = L_{\rm b} \blb$.  Because typical star-disk-binary systems satisfy $L_{\rm b} \gg \Ld, S$, we take $\blb$ to be fixed for the remainder of this work.

\section{Disk Warping}
\label{sec:DiskWarp}

When $\ag \lesssim H/r$ ($\ag$ is the Shakura-Sunyaev viscosity parameter, $H$ is the disk scaleheight), which is satisfied for protostellar disks (e.g. \citealt{Rafikov(2017)}), the main internal torque enforcing disk rigidity and coherent precession comes from bending wave propagation \citep{PapaloizouLin(1995),LubowOgilvie(2000)}.  As bending waves travel at 1/2 of the sound speed, the wave crossing time is of order $t_{\rm bw} = 2(r/H) \Om^{-1}$.  When $t_{\rm bw}$ is longer than the characteristic precession times $\omdb^{-1}$ or $\omds^{-1}$ from an external torque, significant disk warps can be induced.  In the extreme nonlinear regime, disk breaking may be possible \citep{Larwood(1996),Dougan(2015)}. To compare $t_{\rm bw}$ with $\oms^{-1}$ and $\omb^{-1}$, we adopt the disk sound speed profile
\begin{align}
\cs(r) &= H(r) \Om(r) = h_{\rm out} \sqrt{ \frac{G \Mst}{\rout} } \left( \frac{\rout}{r} \right)^q
\nonumber \\
&= h_{\rm in} \sqrt{ \frac{G \Mst}{\rin} } \left( \frac{\rin}{r} \right)^q,
\label{eq:q}
\end{align}
where $h_{\rm in} = H(\rin)/\rin$ and $h_{\rm out} = H(\rout)/\rout$.  Passively heated disks have $q \approx 0.0-0.3$ \citep{ChiangGoldreich(1997)}, while actively heated disks have $q \approx 3/8$ \citep{Lynden-BellPringle(1974)}.  We find
\begin{align}
t_{\rm bw} \oms &= 4.7 \times 10^{-4} \left( \frac{0.1}{h_{\rm in}} \right) \left( \frac{\kq}{0.1} \right) \frac{\bRst^2}{\brin^2} \left( \frac{r}{\rin} \right)^{q-7/2},
\label{eq:tbw_oms}\\
t_{\rm bw} \omb &= 1.7 \times 10^{-2} \left( \frac{0.1}{h_{\rm out}} \right) \frac{\bMb \brout^3}{\bMst \bab^3} \left( \frac{r}{\rout} \right)^{q+3/2}.
\label{eq:tbw_omb}
\end{align}
Thus, we expect the small warp approximation to be valid everywhere in the disk.  This expectation is confirmed by our detailed calculation of disk warps presented later in this section.

Although the disk is flat to a good approximation, the interplay between the disk warp/twist and viscous dissipation can lead to appreciable damping of the misalignment between the disk and the
external perturber (i.e., the oblate star or the binary companion). In particular, when an external torque $\bText$ (per unit mass) is applied to a disk in the bending wave regime (which could be either $\bT_{\rm db}$ or $\bT_{\rm ds}$), the disk's viscosity causes the disk normal to develop a small twist, of order
\be
\frac{\pd \bl}{\pd \ln r} \sim \frac{4\ag}{\cs^2} \bText.
\label{eq:est_twist}
\ee
The detailed derivation of Eq.~\eqref{eq:est_twist} is contained in Sections~\ref{sec:Bin}-\ref{sec:Both}.  Since $\bT_{\rm ext} \propto \bl_{\rm ext} \btimes \bl$ ($\bl_{\rm ext}$ is the axis around which $\bl$ precesses), where the viscous twist interacts with the external torque, effecting the evolution of $\bl$ over viscous timescales.  To an order of magnitude, we have 
\begin{equation}
\bigg| \frac{\der \bl}{\der t} \bigg|_{\rm visc}
\sim  \left\langle   \left( \frac{4\alpha}{\cs^2} \right) \frac{\bText^2}{r^2\Omega} \right\rangle
\sim \left\langle  \frac{4\alpha}{\cs^2} (r^2\Omega)\omega_{\rm ext}^2 \right\rangle,
\end{equation}
where $\omega_{\rm ext}$ is either $\om_{\rm ds}$ or $\om_{\rm db}$, and $\langle \cdots\rangle$ implies proper average over $r$.

We now study the disk warp and viscous evolution quantitatively, using the formalism describing the structure and evolution of circular, weakly warped disks in the bending wave regime.  The relevant equations have been derived by a number of authors \citep{PapaloizouLin(1995),DemianskiIvanov(1997),LubowOgilvie(2000)}.  We choose the formalism of \cite{LubowOgilvie(2000)} and \cite{Lubow(2002)} (see also \citealt{Ogilvie(2006)} when $|\pd \bl / \pd \ln r| \sim 1$), where the evolution of the disk is governed by
\begin{align}
\Sg r^2 \Om \frac{\pd \bl}{\pd t} &= \Sg \bText + \frac{1}{r}\frac{\pd \bG}{\pd r},
\label{eq:dldt_warp} \\
\frac{\pd \bG}{\pd t} &= \left( \frac{\Om^2 - \kappa^2}{2 \Om} \right) \bl \btimes \bG - \ag \Om \bG + \frac{\Sg \cs^2 r^3 \Om}{4} \frac{\pd \bl}{\pd r},
\label{eq:dGdt_warp}
\end{align}
where $\bText$ is the external torque per unit mass acting on the disk, $\kappa = (2 \Om/r) \pd (r^2 \Om)/\pd r|_{z=0}$ is the epicyclic frequency, and $\bG$ is the internal torque, which arises from slightly eccentric fluid particles with velocities sheared around the disk mid-plane \citep{DemianskiIvanov(1997)}.  Eq.~\eqref{eq:dldt_warp} is the 2D momentum equation generalized to non-coplanar disks.  Eq.~\eqref{eq:dGdt_warp} is related to how internal torques generated from disk warps are communicated across the disk under the influence of viscosity and precession from non-Keplarian epicyclic frequencies.  See \cite{NixonKing(2016)} for a qualitative discussion and review of Eqs.~\eqref{eq:dldt_warp}-\eqref{eq:dGdt_warp}.

  We are concerned with two external torques acting on different regions of the disk.  For clarity, we break up our calculations into three subsections, considering disk warps produced by individual torques before examining the combined effects.

\subsection{Disk Warp Induced by Binary Companion}
\label{sec:Bin}

\begin{figure*}
\centering
\includegraphics[scale=0.57]{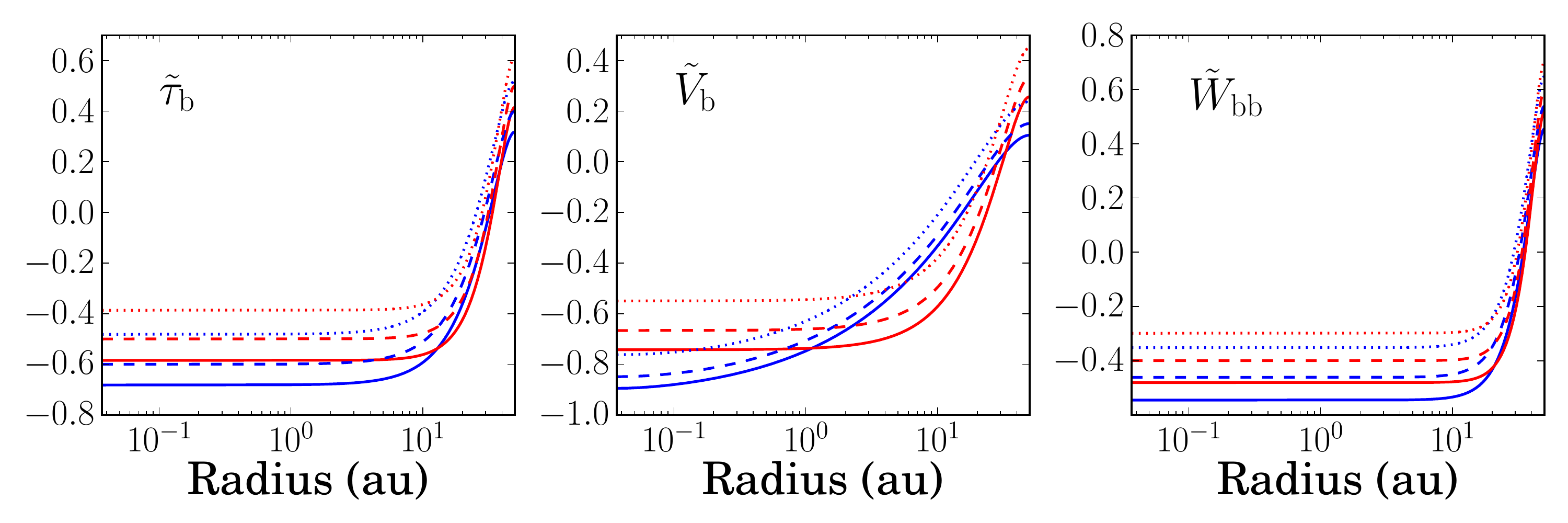}
\caption{
The rescaled radial functions [see Eq.~\eqref{eq:dim_rad} for rescaling] $\tilde \tau_{\rm b}$ [Eq.~\eqref{eq:tb}], $\tilde V_{\rm b}$ [Eq.~\eqref{eq:Vb}], and $\tilde W_{\rm bb}$ [Eq.~\eqref{eq:Wbb}].  We take $(p,q)$ values [Eq.~\eqref{eq:p} and~\eqref{eq:q}] of $p = 0.5$ (solid), $p = 1.0$ (dashed), and $p = 1.5$ (dotted) with $q = 0.0$ (blue) and $q = 0.5$ (red).  All other parameters take their cannonical values [Eq.~\eqref{eq:pars}].   The re-scaled radial functions trace out the viscous twist ($\tilde V_{\rm b}$) and warp ($\tilde \tau_{\rm b}$, $\tilde W_{\rm bb}$) profiles of the disk due to the gravitational torque from the binary companion.
}
\label{fig:rad_bin}
\end{figure*}

\begin{table}
\centering
\begin{tabular}{ |c|c|c|c|c| }
\hline
$p$ & $q$ & $\cUb$ & $\cVb$ & $\cWbb$ \\
\hline
0.5 & 0.5 & 0.857 & 0.857 & 0.857 \\
1.0 & 0.5 & 1.00 & 1.00 & 1.00 \\
1.5 & 0.5 & 1.20 & 1.20 & 1.20 \\
0.5 & 0.0 & 1.65 & 3.67 & 1.32 \\
1.0 & 0.0 & 1.93 & 4.26 & 1.54 \\
1.5 & 0.0 & 2.31 & 5.02 & 1.85  \\
\hline
\end{tabular}
\caption{Dimensionless coefficients $\cUb$ [Eq.~\eqref{eq:tb_scale}], $\cVb$ [Eq.~\eqref{eq:Vb_scale}], and $\cWbb$ [Eq.~\eqref{eq:Wbb_scale}] tabulated for various $p$ and $q$ values [Eqs.~\eqref{eq:p} and~\eqref{eq:q}].  All the parameter values are canonical [Eq.~\eqref{eq:pars}].  When varying $q$, we fix $\hout = 0.1$.
}
\label{tab:bin}
\end{table}

The torque from an external binary companion is given by Eq.~\eqref{eq:Tdb}.  The companion also gives rise to a non-Keplarian epicyclic frequency, given by
\be
\frac{\Om^2-\kappa^2}{2 \Om} = \omb P_2(\bl \bcdot \blb),
\label{eq:kg_bin}
\ee
where $P_l$ are Legendre polynomials.  

To make analytic progress, we take advantage of our expectation that $|\pd \bl/\pd \ln r| \ll 1$ since $t_{\rm bw} \omdb \ll 1$ [see Eq.~\eqref{eq:tbw_omb}].  Specifically, we take
\begin{align}
\bl(r,t) &= \ld(t) + \blone(r,t) + \dots,
\label{eq:l_exp} \\
\bG(r,t) &= \bG_0(r,t) + \bG_1(r,t) + \dots,
\label{eq:G_exp}
\end{align}
where $|\blone| \ll |\ld| = 1$.  Here, $\bG_0(r,t)$ is the internal torque maintaining coplanarity of $\ld(t)$, $\bG_1(r,t)$ is the internal torque maintaining the leading order warp profile $\blone(r,t)$, etc.  To leading order, Eq.~\eqref{eq:dldt_warp} becomes
\be
\Sg r^2 \Om \frac{\der \ld}{\der t} = - \Sg r^2 \Om \omb (\ld \bcdot \blb) \blb \btimes \ld + \frac{1}{r} \frac{\pd \bG_0}{\pd r}.
\label{eq:l0_bin}
\ee
Integrating~\eqref{eq:l0_bin} over $r \der r$, and using the boundary condition
\be
G_0(\rin,t) = G_0(\rout,t) = 0,
\label{eq:G0_bound}
\ee
we obtain
\be
\frac{\der \ld}{\der t} = -\bomb (\ld \bcdot \blb) \blb \btimes \ld,
\label{eq:dl0dt_bin}
\ee
where $\bomb$ is given by
\begin{align}
\bomdb &= \frac{2\pi}{\Ld} \int_{\rin}^{\rout} \omb \Sg r^3 \Om \der r
\nonumber \\
&\simeq \frac{3(5/2-p)}{4(4-p)} \left( \frac{\Mb}{\Ms} \right) \left( \frac{\ab}{\rout} \right)^3 \sqrt{ \frac{G \Ms}{\rout^3} }.
\label{eq:bomdb}
\end{align}
The physical meaning of $\ld$ thus becomes clear: $\ld$ is the unit total angular momentum vector of the disk, or
\be
\ld \equiv \frac{2\pi}{\Ld} \int_{\rin}^{\rout} \Sg r^3 \Om \bl(r,t) \der r.
\label{eq:ld}
\ee
Using Eqs.~\eqref{eq:G0_bound} and~\eqref{eq:dl0dt_bin}, we may solve Eq.~\eqref{eq:l0_bin} for $\bG_0(r,t)$:
\be
\bG_0(r,t) = \gb(r) (\ld \bcdot \blb) \blb \btimes \ld,
\label{eq:G0_bin}
\ee
where
\be
\gb(r) = \int_{\rin}^r (\omb - \bomb) \Sg {r'}^3 \Om \der r'.
\label{eq:gb}
\ee

Using Eqs.~\eqref{eq:G0_bin} and~\eqref{eq:dGdt_warp}, and requiring that $\blone$ not contribute to the total disk angular momentum vector, or
\be
\int_{\rin}^{\rout} \Sg r^3 \Om \blone(r,t) \der r = 0,
\label{eq:l1_cond}
\ee
we obtain the leading order warp $\blone(r,t)$:
\begin{align}
\blone(r,t) = &- \bomb \tb (\ld \bcdot \blb)^2 \blb \btimes (\blb \btimes \ld)
\nonumber \\
&- \Wbb (\ld \bcdot \blb) P_2( \ld \bcdot \blb) \ld \btimes (\blb \btimes \ld)
\nonumber \\
&+ \Vb (\ld \bcdot \blb) \blb \btimes \ld,
\label{eq:l1_bin}
\end{align}
where
\begin{align}
\tb(r) &= \int_{\rin}^r \frac{4 \gb}{\Sg \cs^2 {r'}^3 \Om} \der r' - \tbo,
\label{eq:tb} \\
\Vb(r) &= \int_{\rin}^r \frac{4 \ag \gb}{\Sg \cs^2 {r'}^3} \der r' - \Vbo,
\label{eq:Vb} \\
\Wbb(r) &= \int_{\rin}^r \frac{4 \omb \gb}{\Sg \cs^2 {r'}^3 \Om} \der r' - \Wbbo,
\label{eq:Wbb}
\end{align}
and the constants $X_0$ of the functions $X(r)$ (either $\tb(r)$, $\Vb(r)$, or $\Wbb(r)$) are determined by requiring
\be
\int_{\rin}^{\rout} \Sg r^3 \Om X \der r = 0.
\label{eq:X0}
\ee
Notice the radial functions $\tb$, $\Vb$, and $\Wbb$ trace out the disk's warp profile $|\blone(r)|$ due to the binary companion's gravitational torque.  Because the magnitudes for the radial functions $(2\pi/ \text{Myr})\tb$, $\Vb$, and $\Wbb$ are much smaller than unity everywhere [see Eqs.~\eqref{eq:tb_scale}-\eqref{eq:Wbb_scale}], we define the re-scaled radial function $\tilde X(r) = \ttb, \tVb,$ and $\tWbb$ as
\be
\tilde X(r) \equiv X(r) \Big/ \big[ X(\rout) - X(\rin) \big].
\label{eq:dim_rad}
\ee
Figure~\ref{fig:rad_bin} plots the dimensionless radial functions $\ttb$, $\tVb$, and $\tWbb$ for the canonical parameters of the star-disk-binary system [Eq.~\eqref{eq:pars}].  The scalings of the radial functions evaluated at the outer disk radius are
\begin{align}
 \tb&(\rout) -\tb(\rin) = -1.82 \times 10^{-5} \cUb
\nonumber \\
&\times \left( \frac{0.1}{\hout} \right)^2 \frac{\bMb \brout^{9/2}}{\bMst^{3/2} \bab^6} \frac{\text{Myr}}{2\pi},
\label{eq:tb_scale} \\
\Vb&(\rout) - \Vb(\rin) = -1.54 \times 10^{-3} \cVb
\nonumber \\
&\times \left( \frac{\ag}{0.01} \right) \left( \frac{0.1}{\hout} \right)^2 \frac{\bMb \brout^3}{\bMst \bab^3},
\label{eq:Vb_scale} \\
\Wbb&(\rout) - \Wbb(\rin) = -8.93 \times 10^{-5} \cWbb
\nonumber \\
&\times \left( \frac{0.1}{\hout} \right)^2 \frac{\bMb^2 \brout^6}{\bMst^2 \bab^6}.
\label{eq:Wbb_scale}
\end{align}
Equations~\eqref{eq:tb_scale}-\eqref{eq:Wbb_scale} provide an estimate for the misalignment angle between the disk's inner and outer orbital angular momentum vectors, or $|X(\rout) - X(\rin)| \sim |\bl(\rout,t) \btimes \bl(\rin,t)|$, where $X = (2\pi/\text{Myr})\tb$, $\Vb$, and $\Wbb$.  The dimensionless coefficients $\cUb$, $\cVb$, and $\cWbb$ depend weakly on the parameters $p$, $q$, and $\rin/\rout$.  Table~\ref{tab:bin} tabulates $\cUb$, $\cVb$, and $\cWbb$ for values of $p$ and $q$ as indicated, with the canonical value of $\rin/\rout$ [Eq.~\eqref{eq:pars}].

\subsection{Disk Warp Indued by Oblate Star}
\label{sec:Star}

\begin{figure*}
\centering
\includegraphics[scale=0.57]{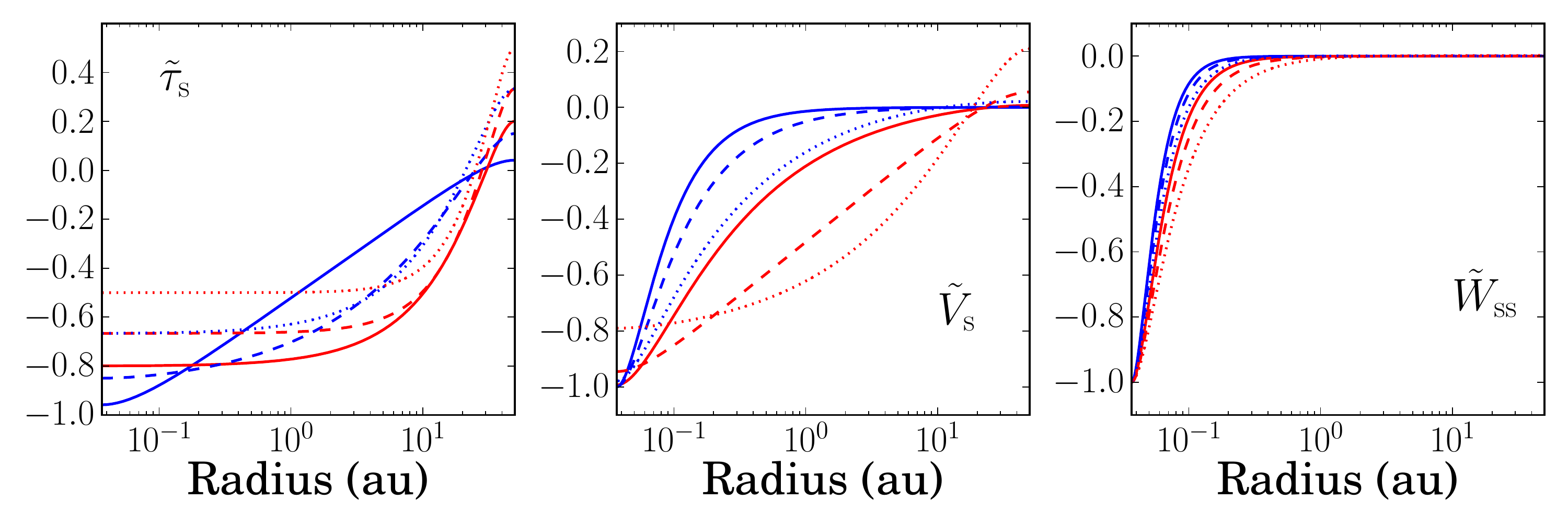}
\caption{
The rescaled radial functions [see Eq.~\eqref{eq:dim_rad} for rescaling] $\tilde \tau_{\rm s}$ [Eq.~\eqref{eq:ts}], $\tilde V_{\rm s}$ [Eq.~\eqref{eq:Vs}], and $\tilde W_{\rm ss}$ [Eq.~\eqref{eq:Wss}].  We take $(p,q)$ values [Eq.~\eqref{eq:p} and~\eqref{eq:q}] of $p = 0.5$ (solid), $p = 1.0$ (dashed), and $p = 1.5$ (dotted) with $q = 0.0$ (blue) and $q = 0.5$ (red).  Other parameters assume their canonical values [Eq.~\eqref{eq:pars}].  The re-scaled radial functions trace out the viscous twist ($\tilde V_{\rm s}$) and warp ($\tilde \tau_{\rm s}$, $\tilde W_{\rm ss}$) profiles of the disk due to the gravitational torque from the oblate star.
}
\label{fig:rad_star}
\end{figure*}

\begin{table}
\centering
\begin{tabular}{ |c|c|c|c|c| }
\hline
$p$ & $q$ & $\cUs$ & $\cVs$ & $\cWss$ \\
\hline
0.5 & 0.5 & 2.66 & 0.315 & 0.800 \\
1.0 & 0.5 & 1.00 & 1.00 & 1.00 \\
1.5 & 0.5 & 0.400 & 6.18 & 1.33 \\
0.5 & 0.0 & 24.2 & 0.0735 & 0.457 \\
1.0 & 0.0 & 4.28 & 0.110 & 0.533 \\
1.5 & 0.0 & 1.20 & 0.207 & 0.639 \\
\hline
\end{tabular}
\caption{Dimensionless coefficients $\cUs$ [Eq.~\eqref{eq:ts_scale}], $\cVs$ [Eq.~\eqref{eq:Vs_scale}], and $\cWss$ [Eq.~\eqref{eq:Wss_scale}], for different values of $p$ and $q$ [Eqs.~\eqref{eq:p} and~\eqref{eq:q}].  All other parameter values are canonical [Eq.~\eqref{eq:pars}].  When varying $q$, we fix $\hout = 0.1$.  }
\label{tab:star}
\end{table}

The torque on the disk from the oblate star is given by Eq.~\eqref{eq:Tds}.  The stellar quadrupole moment also gives rise to a non-Keplarian epicyclic frequency given by
\be
\frac{\Om^2 - \kappa^2}{2 \Om} = \oms P_2(\bl \bcdot \bs).
\label{eq:kg_star}
\ee
Equations~\eqref{eq:dldt_warp}-\eqref{eq:dGdt_warp} are coupled with the motion of the host star's spin axis:
\be
S \frac{\der \bs}{\der t} = - \int_{\rin}^{\rout} \Big[ 2\pi \Sg r^3 \Om \oms (\bs \bcdot \bl) \bl \btimes \bs \Big] \der r,
\label{eq:dsdt_warp_star}
\ee
Expanding $\bl$ and $\bG$ according to Eqs.~\eqref{eq:l_exp} and~\eqref{eq:G_exp}, integrating Eq.~\eqref{eq:dldt_warp} over $r \der r$, and using the boundary condition~\eqref{eq:G0_bound}, we obtain the leading order evolution equations
\begin{align}
\frac{\der \bs}{\der t} &= - \bomd (\bs \bcdot \ld) \ld \btimes \bs, \\
\frac{\der \ld}{\der t} &= - \boms (\ld \bcdot \bs) \bs \btimes \ld,
\end{align}
where (assuming $\rin \ll \rout$)
\begin{align}
\bomds &= \frac{2\pi}{\Ld} \int_{\rin}^{\rout} \omds \Sg r^3 \Om \der r
\nonumber \\
&\simeq \frac{3(5/2-p)\kq}{2(1+p)} \frac{\Rs^2 \bOmst^2}{\rout^{1-p} \rin^{1+p}} \sqrt{ \frac{G \Ms}{\rout^3} },
\label{eq:bomds} \\
\bomsd &= (\Ld/S)\bomds
\nonumber \\
&\simeq \frac{3(2-p)\kq}{2(1+p)\ks} \left( \frac{\Md}{\Ms} \right) \bOmst \frac{\sqrt{G \Ms \Rs^3}}{\rout^{2-p} \rin^{1+p}}.
\label{eq:bomsd}
\end{align}
With $\der \ld/\der t$ and $\der \bs/\der t$ determined, Eq.~\eqref{eq:dldt_warp} may be integrated to obtain the leading order internal torque:
\be
\bG_0(r,t) = \gs(r) (\ld \bcdot \bs) \bs \btimes \ld,
\ee
where
\be
\gs(r) = \int_{\rin}^r (\oms - \boms) \Sg {r'}^3 \Om \der r'.
\ee
Similarly, the leading order warp profile is
\begin{align}
\blone(r,t) = &- \bomd \ts (\ld \bcdot \bs)^2 (\ld \btimes \bs) \btimes \ld
\nonumber \\
&- \boms \ts (\ld \bcdot \bs)^2 \bs \btimes (\bs \btimes \ld)
\nonumber \\
&- \Wss (\ld \bcdot \bs) P_2( \ld \bcdot \bs) \ld \btimes (\bs \btimes \ld)
\nonumber \\
&+ \Vs (\ld \bcdot \bs) \bs \btimes \ld,
\label{eq:l1_star}
\end{align}
where
\begin{align}
\ts(r) &= \int_{\rin}^r \frac{4 \gs}{\Sg \cs^2 {r'}^3 \Om} \der r' - \tso,
\label{eq:ts} \\
\Vs(r) &= \int_{\rin}^r \frac{4 \ag \gs}{\Sg \cs^2 {r'}^3} \der r' - \Vso,
\label{eq:Vs} \\
\Wss(r) &= \int_{\rin}^r \frac{4 \oms \gs}{\Sg \cs^2 {r'}^3 \Om} \der r' - \Wsso.
\label{eq:Wss}
\end{align}
In Figure~\ref{fig:rad_star}, we plot the rescaled radial functions $\tts$, $\tVs$, and $\tWss$ for various $p$ and $q$ values, tracing out the re-scaled warp profile across the radial extent of the disk due to the oblate star's torque.  The radial function differences evaluated at the disk's outer and inner truncation radii are
\begin{align}
\ts&(\rout) - \ts(\rin)  = 2.21 \times 10^{-6}\cUs  \left( \frac{0.1}{\hout} \right)^2 \left( \frac{\kq}{0.1} \right)
\nonumber \\
&\times  \left( \frac{1358 \, \brout}{\brin} \right)^{p-1} \frac{\bRst^2 \brout^{3/2}}{\brin^{2} \bMst^{1/2}} \left( \frac{\bOmst}{0.1} \right)^2 \frac{\text{Myr}}{2\pi},
\label{eq:ts_scale} \\
\Vs&(\rout)-\Vs(\rin) = 1.13 \times 10^{-3} \cVs
\nonumber \\
 &\times \left( \frac{\ag}{0.01} \right) \left( \frac{0.1}{\hin} \right)^2 \left(\frac{\kq}{0.1} \right) 
\frac{\bRst^2}{\brin^2} \left( \frac{\bOmst}{0.1} \right)^2,
\label{eq:Vs_scale} \\
\Wss&(\rout)-\Wss(\rin) = 4.39 \times 10^{-7} \cWss 
\nonumber \\
&\times \left( \frac{\kq}{0.1} \right)^2 \left( \frac{0.1}{\hin} \right)^2 \frac{\bRst^4}{\brin^4} \left( \frac{\bOmst}{0.1} \right)^2.
\label{eq:Wss_scale}
\end{align}
Equations~\eqref{eq:ts_scale}-\eqref{eq:Wss_scale} provide an estimate for the misalignment angle between the disk's outer and inner orbital angular momentum unit vectors $|\bl(\rout,t) \btimes \bl(\rin,t)|$ due to the oblate star's torque.  The dimensionless coefficients $\cUs$, $\cVs$, and $\cWss$ depend weakly on the parameters $p$, $q$, and $\rin/\rout$.  In Table~\ref{tab:star}, we tabulate $\cUs$, $\cVs$, and $\cWss$ for the $p$ and $q$ values indicated, with $\rin/\rout$ taking the canonical value [Eq.~\eqref{eq:pars}].

\subsection{Disk Warps Induced by Combined Torques}
\label{sec:Both}

\begin{figure*}
\centering
\includegraphics[scale=0.57]{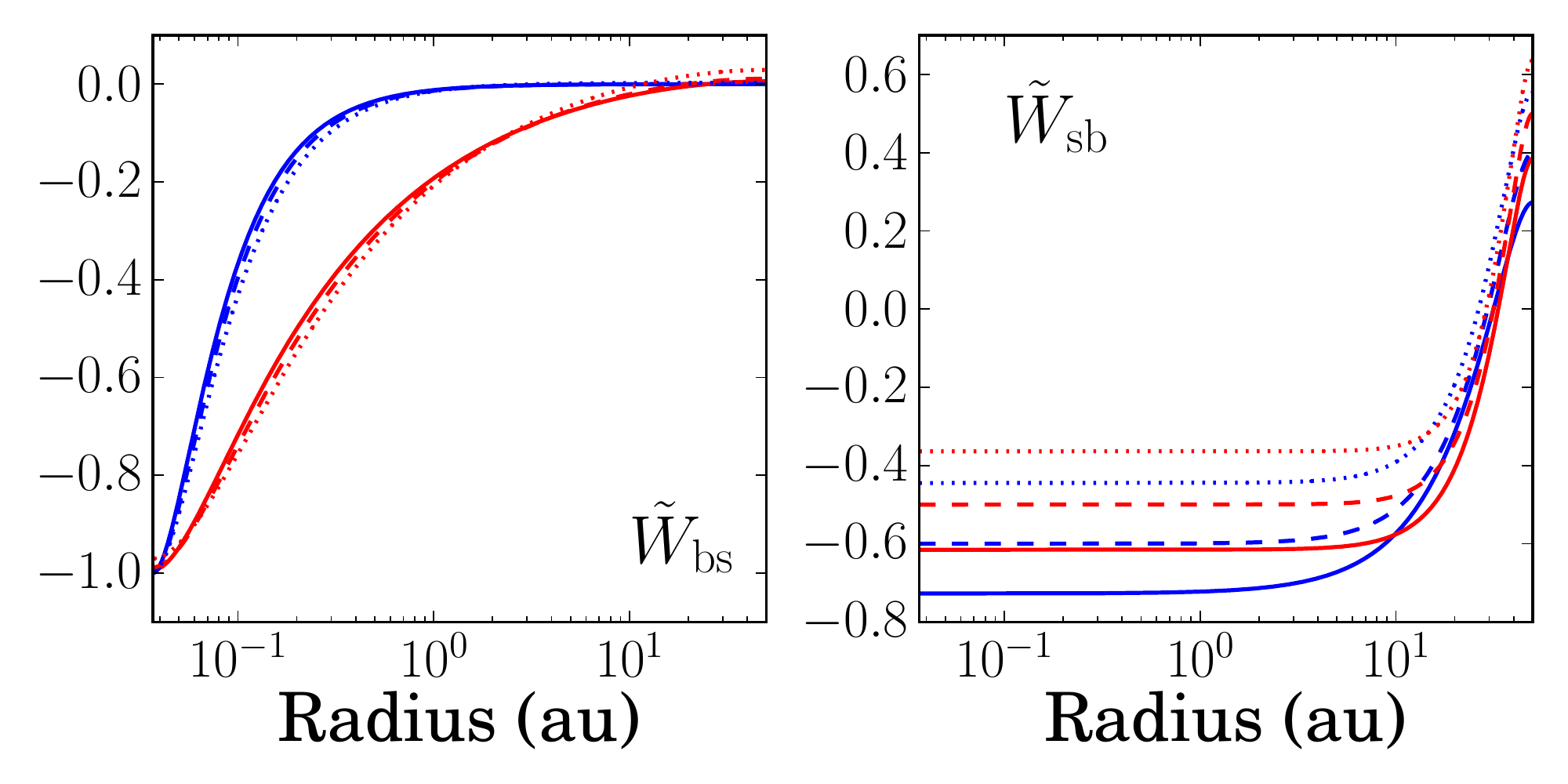}
\caption{
The rescaled radial functions [see Eq.~\eqref{eq:dim_rad} for rescaling] $\tilde W_{\rm bs}$ [Eq.~\eqref{eq:Wbs}], and $\tilde W_{\rm sb}$ [Eq.~\eqref{eq:Wsb}].  We take $(p,q)$ values [Eq.~\eqref{eq:p} and~\eqref{eq:q}] of $p = 0.5$ (solid), $p = 1.0$ (dashed), and $p = 1.5$ (dotted) with $q = 0.0$ (blue) and $q = 0.5$ (red).  We take all parameters to be cannonical [Eq.~\eqref{eq:pars}].  The re-scaled radial functions trace out the the warp ($\tilde W_{\rm bs}$, $\tilde W_{\rm sb}$) profiles of the disk due to the interaction between the binary companion and oblate star torques (see text for discussion).
}
\label{fig:rad_both}
\end{figure*}

\begin{table}
\centering
\begin{tabular}{ |c|c|c|c|}
\hline
$p$ & $q$ & $\cWbs$ & $\cWsb$ \\
\hline
0.5 & 0.5 & 2.13 & 0.917 \\
1.0 & 0.5 & 1.00 & 1.00 \\
1.5 & 0.5 & 0.457 & 1.06 \\
0.5 & 0.0 & 4.57 & 312 \\
1.0 & 0.0 & 1.93 & 319 \\
1.5 & 0.0 & 0.823 & 307 \\
\hline
\end{tabular}
\caption{Dimensionless coefficients $\cWbs$ [Eq.~\eqref{eq:Wbs_scale}] and $\cWsb$ [Eq.~\eqref{eq:Wsb_scale}] for values of $p$ and $q$ as indicated [Eqs.~\eqref{eq:p} and~\eqref{eq:q}].  All parameter values are canonical [Eq.~\eqref{eq:pars}].  When varying $q$, we fix $\hout = 0.1$.  }
\label{tab:both}
\end{table}

The combined torques from the distant binary and oblate star are given by Eqs.~\eqref{eq:Tdb} and~\eqref{eq:Tds}, and the non-Keplarian epicyclic frequencies are given by Eqs.~\eqref{eq:kg_bin} and~\eqref{eq:kg_star}.  Using the same procedure as Sections~\ref{sec:Bin}-\ref{sec:Star}, the leading order correction to the disk's warp is
\begin{align}
\blone(r,t) = \, & (\blone)_\text{bin} + (\blone)_\text{star}
\nonumber \\
&- \boms \tb (\ld \cdot \bs) \Big[ (\bs \btimes \ld) \bcdot \blb \Big] \blb \btimes \ld
\nonumber \\
&- \boms \tb (\ld \bcdot \blb)(\ld \bcdot \bs) \blb \btimes (\bs \btimes \ld)
\nonumber \\
&- \bomb \ts (\ld \bcdot \blb) \Big[ (\blb \btimes \ld) \bcdot \bs \Big] \bs \btimes \ld
\nonumber \\
&- \bomb \ts (\ld \bcdot \bs) (\ld \bcdot \blb) \bs \btimes (\blb \btimes \ld)
\nonumber \\
&-\Wsb (\ld \bcdot \blb) P_2(\ld \bcdot \bs) \ld \btimes (\blb \btimes \ld)
\nonumber \\
&-\Wbs (\ld \bcdot \bs) P_2(\ld \bcdot \blb) \ld \btimes (\bs \btimes \ld),
\label{eq:l1_both}
\end{align}
where $(\blone)_\text{bin}$ is Eq.~\eqref{eq:l1_bin}, $(\blone)_\text{star}$ is Eq.~\eqref{eq:l1_star}, $\tb$ and $\ts$ are given in Eqs.~\eqref{eq:tb} and~\eqref{eq:ts}, and 
\begin{align}
\Wbs(r) &= \int_{\rin}^r \frac{4 \omb \gs}{\Sg \cs^2 {r'}^3 \Om} \der r' - \Wbso,
\label{eq:Wbs} \\
\Wsb(r) &= \int_{\rin}^r \frac{4 \oms \gb}{\Sg \cs^2 {r'}^3 \Om} \der r' - \Wsbo.
\label{eq:Wsb}
\end{align}
Notice $\blone$ is not simply the sum $\blone = (\blone)_{\rm bin} + (\blone)_{\rm star}$.  The cross $\omds \tb$ ($\omdb \ts$) terms come from the motion of the internal torque resisting $\bTs$ ($\bTb$) induced by $\bTb$ ($\bTs$).  The cross $\Wbs$ ($\Wsb$) terms come from the internal torque resisting $\bTs$ ($\bTb$) twisted by the non-Keplarian epicyclic frequency induced by the binary [Eq.~\eqref{eq:kg_bin}] [star, Eq.~\eqref{eq:kg_star}].  In Figure~\ref{fig:rad_both}, we plot the re-scaled radial functions $\tWbs$ and $\tWsb$ for various $p$ and $q$ values, tracing out the warp profile across the radial extent of the disk due to the combined binary and stellar torques.  The radial functions $\Wbs$ and $\Wsb$ evaluated at the disk's outer and inner truncation radii are
\begin{align}
\Wbs&(\rout)-\Wbs(\rin) = - 7.23 \times 10^{-6} \cWbs \left( \frac{0.1}{\hout} \right)^2 
\nonumber \\
&\times   \left( \frac{\kq}{0.1} \right) \left( \frac{1358 \, \brout}{\brin} \right)^{p-1} \frac{\bMb \bRst^2 \brout^3}{\bMst \bab^3 \brin^2} \left( \frac{\bOmst}{0.1} \right)^2,
\label{eq:Wbs_scale} \\
\Wsb&(\rout)-\Wsb(\rin) = 1.23 \times 10^{-9} \cWsb
\nonumber \\
&\times \left( \frac{0.1}{\hout} \right)^2 \left( \frac{\kq}{0.1} \right) \frac{\bMb \bRst \brout}{\bMst \bab^3} \left( \frac{\bOmst}{0.1} \right)^2.
\label{eq:Wsb_scale}
\end{align}
 These provide an estimate for the misalignment angle between the disk's outer and inner orbital angular momentum unit vectors $|\bl(\rout,t) \btimes \bl(\rin,t)|$ due to the binary and stellar torques.  The dimensionless coefficients $\cWbs$ and $\cWsb$ depend on the parameters $p$, $q$, and $\rin/\rout$.  Table~\ref{tab:both} tabulates $\cWbs$ and $\cWsb$ for several $p$ and $q$ values, with $\rin/\rout$ taking the canonical value [Eq.~\eqref{eq:pars}].

\subsection{Disk Warp Profile: Summary}
\label{sec:Warp_Summary}

\begin{figure}
\centering
\includegraphics[scale=0.5]{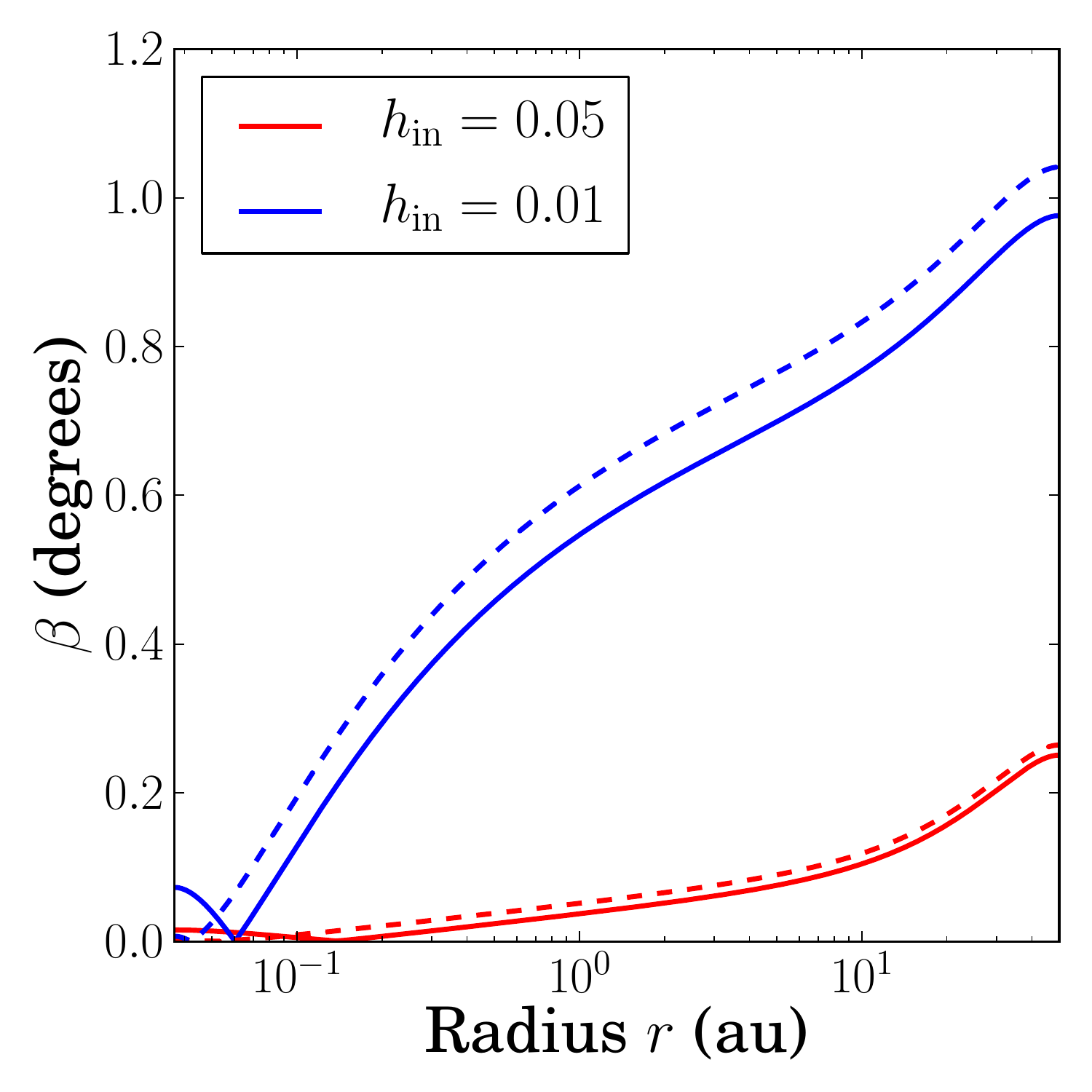}
\caption{ 
Disk misalignment angle $\bg$ [Eq.~\eqref{eq:bg}] as a function of radius $r$, for the $h_{\rm in}$ [Eq.~\eqref{eq:q}] values indicated, all for $h_{\rm out} = 0.05$ [Eq.~\eqref{eq:q}].  The disk masses are $\Md = 0.1 \, \Msun$ (solid) and $\Md = 0.01 \, \Msun$ (dashed), with $p = 1$ [Eq.~\eqref{eq:p}], $\ag = 0.01$, $\ab = 300 \, \text{au}$, and $\bs$, $\ld$, and $\blb$ lying in the same plane with $\tgsd = \tgdb = 30^\circ$.
}
\label{fig:profile_far}
\end{figure}

\begin{figure}
\centering
\includegraphics[scale=0.5]{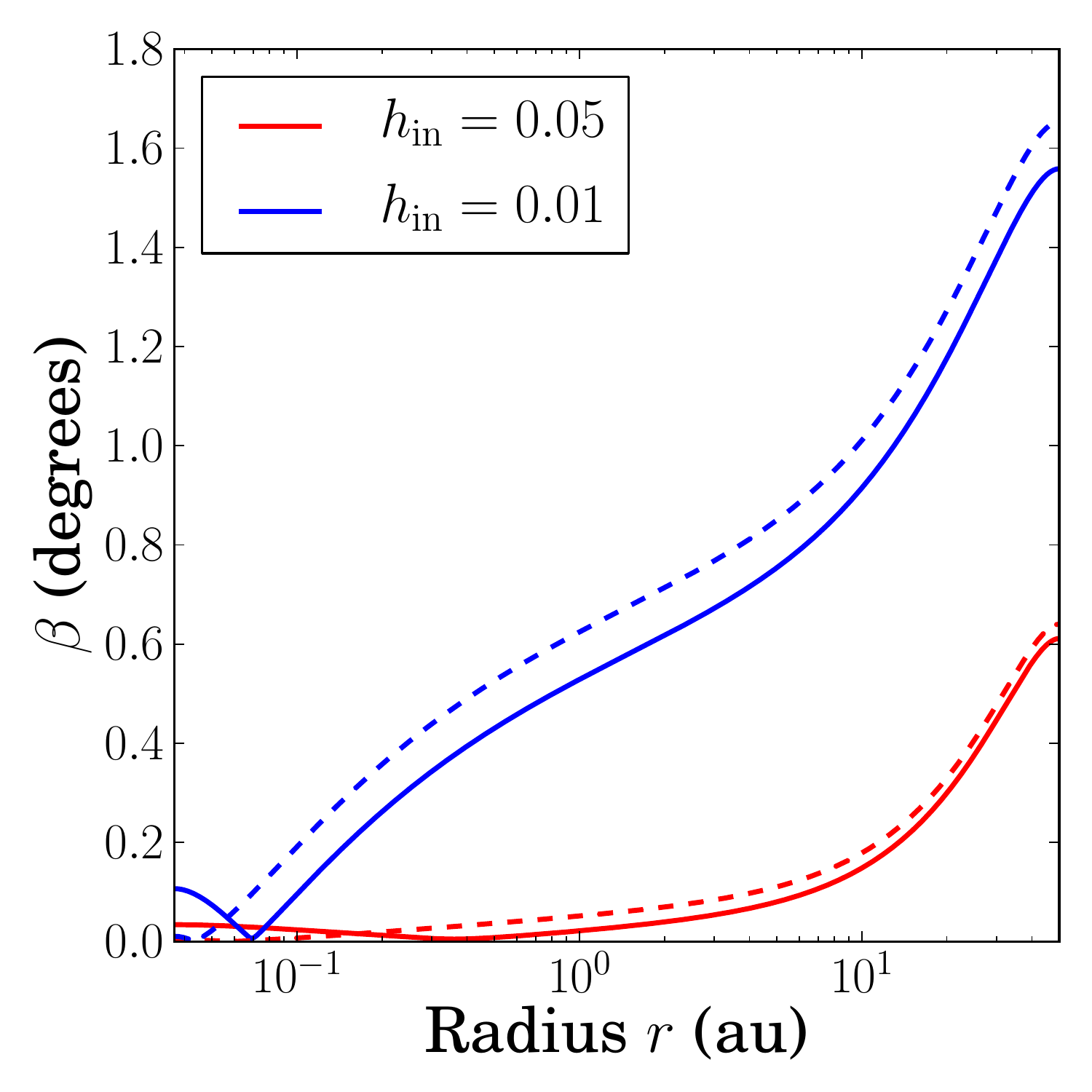}
\caption{ 
Same as Fig.~\protect\ref{fig:profile_far}, except $\ab = 200 \, \text{au}$.
}
\label{fig:profile_close}
\end{figure}

In the previous subsections, we have derived semi-analytic expressions for the disk warp profiles due to the combined torques from the oblate host star and the binary companion.  Our general conclusion is that the warp is quite small across the whole disk. We illustrate this conclusion with a few examples (Figs.~\ref{fig:profile_far}-\ref{fig:profile_close}).   We define the disk misalignment angle $\bg = \bg(r,t)$ as the misalignment of the disk's local angular momentum unit vector $\bl(r,t)$ by
\be
\sin \bg(r,t) \equiv \big| \bl(r,t) \btimes \ld(t) \big|,
\label{eq:bg}
\ee
where $\ld$ is unit vector along the total angular momentum of the disk [Eq.~\eqref{eq:ld}]. 

Figures~\ref{fig:profile_far}-\ref{fig:profile_close} that the disk warp angle is less than a few degrees for the range of parameters considered.  When $\hin = 0.05$, the binary's torque has the strongest influence on the disk's warp profile.  As a result, the disk warp ($\pd \bg/\pd \ln r$) is strongest near the disk's outer truncation radius ($r \gtrsim 10 \, \text{au}$).  When $\hin = 0.01$, the spinning star's torque has a strong influence on the disk's warp profile, and the warp becomes large near the inner truncation radius ($r \lesssim 1 \, \text{au}$).

Notice that the differences between the high disk-mass ($\Md = 0.1 \, \Msun$, solid lines) and low disk-mass ($\Md = 0.01 \Msun$, dashed lines) are marginal.  This is because only the precession rate of the star around the disk $\bomsd$ [Eq.~\eqref{eq:bomsd}] depends on the disk mass, and it enters the disk warp profile only through the term $\bomsd \ts$ [see Eq.~\eqref{eq:l1_star}].  Because the disk's internal torque from bending waves is purely hydrodynamical, the other terms in the disk warp profile are independent of the disk mass.

\subsection{Viscous Evolution}
\label{sec:Visc}

\begin{table}
\centering
\begin{tabular}{ |c|c|c|c|c| }
\hline
$p$ & $q$ & $\Gamma_{\rm b}$ & $\Gamma_{\rm s}$ & $\Gamma_{({\rm bs})}$ \\
\hline
0.5 & 0.5 & 0.698 & 0.522 & 1.70 \\
1.0 & 0.5 & 1.00 & 1.00 & 1.00 \\
1.5 & 0.5 & 1.41 & 2.86 & 0.527 \\
0.5 & 0.0 & 1.44 & 0.0964 & 8.64 \\
1.0 & 0.0 & 2.31 & 0.0970 & 5.38 \\
1.5 & 0.0 & 3.82 & 0.108 & 3.05 \\
\hline
\end{tabular}
\caption{Dimensionless viscosity coefficients $\Gamma_{\rm b}$ [Eq.~\eqref{eq:cgb}], $\Gamma_{\rm s}$ [Eq.~\eqref{eq:cgs}], and $\Gamma_{({\rm bs})}$ [Eq.~\eqref{eq:cgbs}], for various $p$ and $q$ values.  All other parameter values are canonical [Eq.~\eqref{eq:pars}].  }
\label{tab:visc}
\end{table}

\begin{figure}
\centering
\includegraphics[scale=0.65]{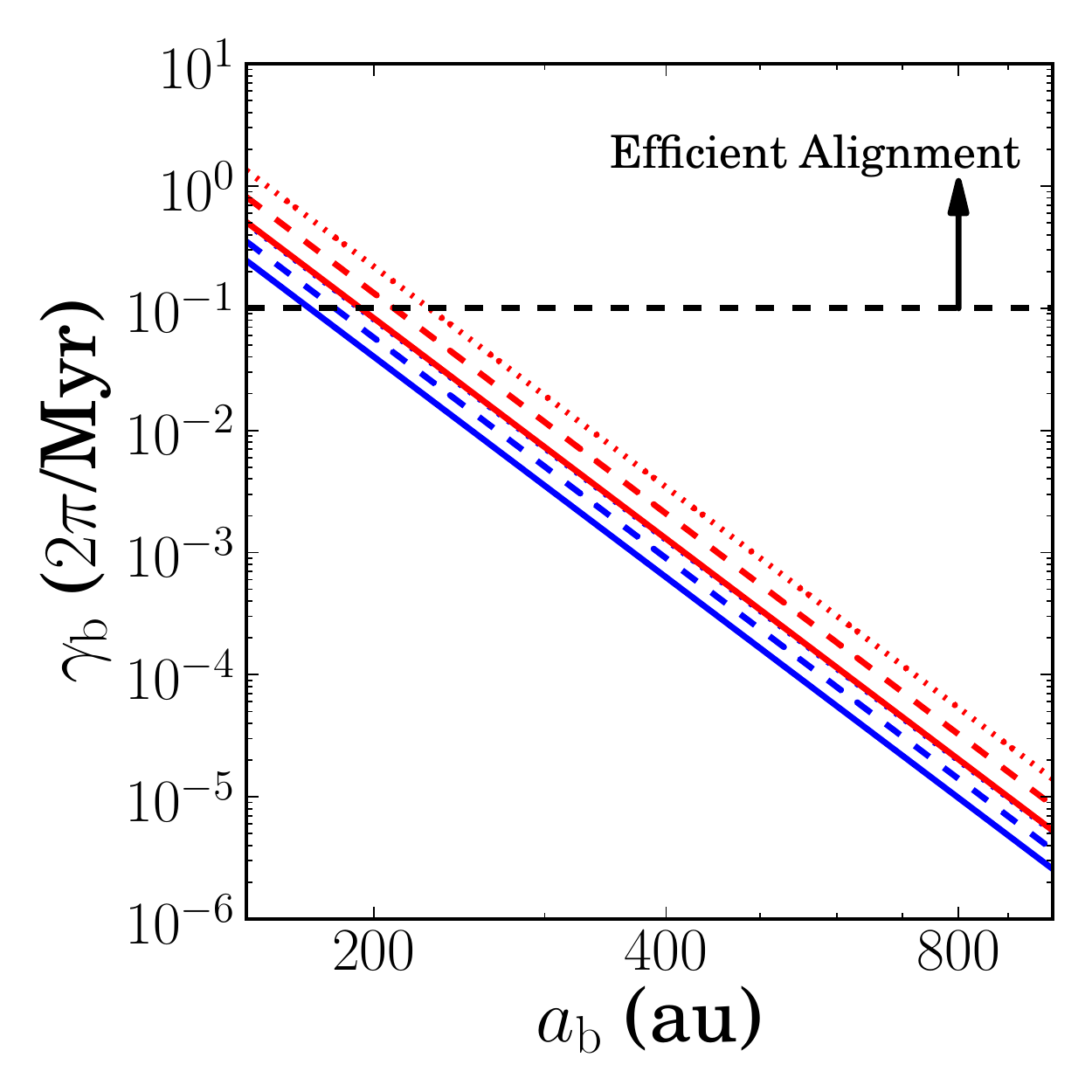}
\caption{ 
The damping rate $\cgb$ [Eq.~\eqref{eq:cgb}] as a function of the binary semi-major axis $\ab$.    We take the $p$ [Eq.~\eqref{eq:p}] value to be $p = 0.5$ (solid), $p = 1.0$ (dashed), and $p = 1.5$ (dotted), with the $q$ [Eq.~\eqref{eq:q}] value of $q = 0.0$ (blue) and $q = 0.5$ (red).  We take all other parameter values to be canonical [Eq.~\eqref{eq:pars}].  When varying $q$, we fix $h_{\rm out} = 0.05$ [Eq.~\eqref{eq:q}].  When the damping rate $\cgb \gtrsim 0.1 (2\pi/\text{Myr})$, viscous torques from disk warping may significantly decrease the mutual disk-binary inclination $\tg_{\rm db}$ [Eq.~\eqref{eq:tgdb}] over the disk's lifetime.
}
\label{fig:cgb}
\end{figure}

\begin{figure}
\centering
\includegraphics[scale=0.65]{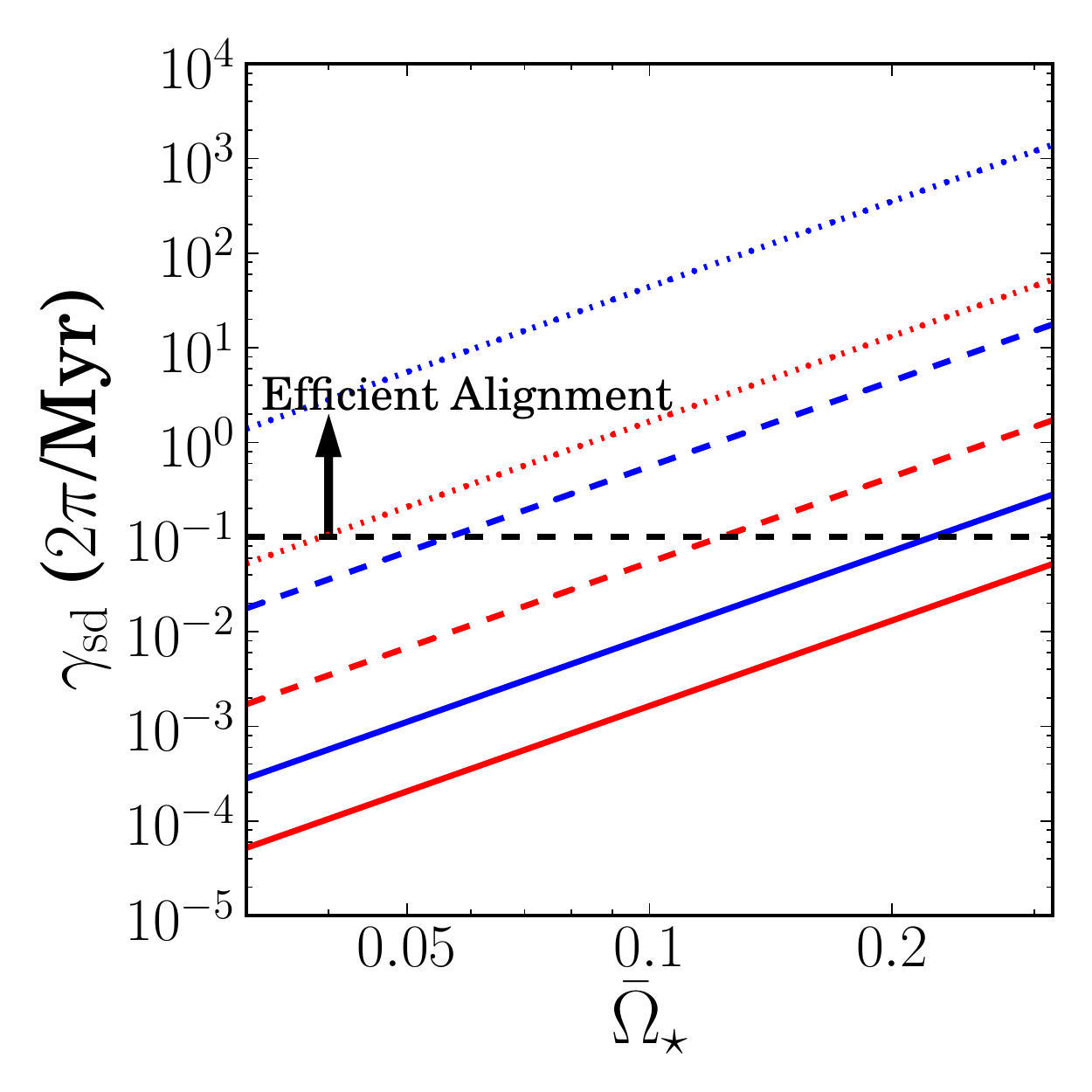}
\caption{ 
The damping rate $\cg_{\rm sd}$ [Eq.~\eqref{eq:cgsd}] as a function of the normalized stellar rotation frequency $\bOmst$ [Eq.~\eqref{eq:pars}].  We take the $p$ [Eq.~\eqref{eq:p}] values to be $p = 0.5$ (solid), $p = 1.0$ (dashed), and $p = 1.5$ (dotted), with $q$ [Eq.~\eqref{eq:q}] values of $q = 0.0$ (blue) and $q = 0.5$ (red).  We take all other parameter values to be canonical [Eq.~\eqref{eq:pars}].  When varying $q$, we fix $h_{\rm in} = 0.03$ [Eq.~\eqref{eq:q}].  When the damping rate $\cg_{\rm sd} \gtrsim 0.1 (2\pi/\text{Myr})$, viscous torques from disk warping may significantly decrease the mutual star-disk inclination $\tg_{\rm sd}$ [Eq.~\eqref{eq:tgsd}] over the disk's lifetime.
}
\label{fig:cgsd}
\end{figure}

As noted above, when a hydrodynamical disk in the bending wave regime is torqued externally, viscosity causes the disk to develop a small twist, which exerts a back-reaction torque on the disk.  When torqued by a central oblate star and a distant binary, the leading order viscous twist in the disk is
\be
(\blone)_{\rm visc} = \Vb (\blb \bcdot \ld) \blb \btimes \ld + \Vs (\bs \bcdot \ld) \bs \btimes \ld,
\ee
where $\Vb$ and $\Vs$ are defined in Eqs.~\eqref{eq:Vb} and~\eqref{eq:Vs}.  All other terms in Eq.~\eqref{eq:l1_both} are non-dissipative, and do not contribute to the alignment evolution of the disk.  Inserting $(\blone)_{\rm visc}$ into Eqs.~\eqref{eq:dldt_warp} and~\eqref{eq:dsdt_warp_star}, and integrating over $2 \pi r \der r$, we obtain
\begin{align}
\left( \frac{\der \bLd}{\der t} \right)_{\rm visc} = \, &\Ld \cgb (\ld \bcdot \blb)^2 \blb \btimes (\blb \btimes \ld)
\nonumber \\
&+ \Ld \cgs (\ld \bcdot \bs)^2 \bs \btimes (\bs \btimes \ld) 
\nonumber \\
&+ \Ld \cgbs (\ld \bcdot \blb)(\ld \bcdot \bs)  \blb \btimes (\bs \btimes \ld) 
\nonumber \\
 &+ \Ld \cgbs (\ld \bcdot \bs)(\ld \bcdot \blb) \bs \btimes (\blb \btimes \ld) ,
\label{eq:dbLddt_visc} \\
\left( \frac{\der {\bm S}}{\der t} \right)_{\rm visc} = \, &-\Ld \cgs (\ld \bcdot \bs)^2 \bs \btimes (\bs \btimes \ld)
\nonumber \\
 &- \Ld \cgbs (\ld \bcdot \bs)(\ld \bcdot \blb) \bs \btimes (\blb \btimes \ld),
 \label{eq:dbSdt_visc}
\end{align}
where
 \begin{align}
 \cgb &\equiv \frac{2\pi}{\Ld} \int_{\rin}^{\rout} \frac{4 \ag \gb^2}{\Sg \cs^2 r^3} \der r 
 \nonumber \\
 &= - \frac{2\pi}{\Ld}\int_{\rin}^{\rout} \Sg r^3 \Om (\omb - \bomb) \Vb \der r, 
 \label{eq:cgb_app} \\
 \cgs &\equiv \frac{2\pi}{\Ld} \int_{\rin}^{\rout} \frac{4 \ag \gs^2}{\Sg \cs^2 r^3} \der r 
 \nonumber \\
 &= -\frac{2\pi}{\Ld} \int_{\rin}^{\rout} \Sg r^3 \Om (\oms - \boms) \Vs \der r,
 \label{eq:cgs_app} \\
 \cgbs &\equiv \frac{2\pi}{\Ld} \int_{\rin}^{\rout} \frac{4 \ag \gb \gs}{\Sg \cs^2 r^3} \der r 
 \nonumber \\
 &= -\frac{2\pi}{\Ld} \int_{\rin}^{\rout} \Sg r^3 \Om (\oms - \boms) \Vb \der r 
 \nonumber \\
 &= - \frac{2\pi}{\Ld} \int_{\rin}^{\rout} \Sg r^3 \Om (\omb - \bomb) \Vs \der r .
 \label{eq:cgbs_app}
 \end{align}
When deriving Eqs.~\eqref{eq:dbLddt_visc} and~\eqref{eq:dbSdt_visc}, we have neglected terms proportional to $\blone \bcdot \bs$ or $\blone \bcdot \blb$, as these only modify the dynamics by changing the star-disk and disk-binary precessional frequencies, respectively.  Using
\begin{align}
\frac{\der \ld}{\der t} &= \frac{1}{\Ld} \left( \frac{\der \bLd}{\der t} - \ld \frac{\der \Ld}{\der t} \right), \\
\frac{\der \bs}{\der t} &= \frac{1}{S} \left( \frac{\der \bS}{\der t} - \bs \frac{\der S}{\der t} \right),
\end{align}
the leading order effect of viscous disk twisting on the time evolution of $\ld$ and $\bs$ is
\begin{align}
\bigg( \frac{\der \ld}{\der t} \bigg)_{\rm visc} = \, &\cgb (\ld \bcdot \blb)^3 \ld \btimes (\blb \btimes \ld)
\nonumber \\
&+ \cgs (\ld \bcdot \bs)^3 \ld \btimes (\bs \btimes \ld)
\nonumber \\
&+ \cgbs (\ld \bcdot \blb)(\ld \bcdot \bs)^2 \ld \btimes (\blb \btimes \ld)
\nonumber \\
&+ \cgbs (\ld \bcdot \bs) (\ld \bcdot \blb)^2 \ld \btimes (\bs \btimes \ld),
\label{eq:dldt_visc} \\
\bigg( \frac{\der \bs}{\der t} \bigg)_{\rm visc} = \, & -\frac{\Ld}{S} \cgs (\ld \bcdot \bs)^2 \bs \btimes (\bs \btimes \ld)
\nonumber \\
 &- \frac{\Ld}{S} \cgbs (\ld \bcdot \bs)(\ld \bcdot \blb) \bs \btimes (\blb \btimes \ld) .
 \label{eq:dsdt_visc}
 \end{align}
 
 The four terms in $(\der \ld/\der t)_{\rm visc}$ [Eq.~\eqref{eq:dldt_visc}] arises from four different back-reaction torques of the disk in response to $\bTs$ [Eq.~\eqref{eq:Tds}] and $\bTb$ [Eq.~\eqref{eq:Tdb}].  To resist the influence of the two external torques $\bTs$ and $\bTb$,  the disk develops two twists $(\pd \bl/\pd \ln r)_{\rm ds}$ and $(\pd \bl/\pd \ln r)_{\rm db}$, given by Eqs.~\eqref{eq:l1_star} and~\eqref{eq:l1_bin}.  The terms in Eqs.~\eqref{eq:dldt_visc}-\eqref{eq:dsdt_visc} proportional to $\cgs$ arise from the back reaction of $(\pd \bl/\pd \ln r)_{\rm ds}$ to $\bTs$, and works to align $\bs$ with $\ld$.  The term in Eq.~\eqref{eq:dldt_visc} proportional to $\cgb$ arises from the back reaction of $\bTb$ to $(\pd \bl/ \pd \ln r)_{\rm db}$ , and works to align $\ld$ with $\blb$.  Because $\cgbs < 0$, the terms in Eqs.~\eqref{eq:dldt_visc}-\eqref{eq:dsdt_visc} proportional to $\cgbs$ have different effects than the terms proportional to $\cgs$ and $\cgb$.  One of the terms in Eqs.~\eqref{eq:dldt_visc}-\eqref{eq:dsdt_visc} proportional to $\cgbs$ arises from the back reaction of $\bTs$ to $(\pd \bl/\pd \ln r)_{\rm db}$, and works to drive $\ld$ perpendicular to $\bs$, while the other arises from the back-reaction of $\bTb$ to $(\pd \bl/\pd \ln r)_{\rm ds}$, and works to drive $\ld$ perpendicular to $\blb$.  Although typically $|\cgs| > |\cgbs|$ or $|\cgb| > |\cgbs|$ (so the dynamical effect of $\cgbs$ may be absorbed into $\cgb$ and $\cgs$), the magnitude of $\cgbs$ is not negligible compared to $\cgs$ and $\cgb$.  For completeness, we include the effects of the $\cgbs$ terms in the analysis below.

The damping rates~\eqref{eq:cgb_app}-\eqref{eq:cgbs_app} may be evaluated and rescaled to give
\begin{align}
\cgb = \, &1.26 \times 10^{-9} \Gamma_{\rm b}\left( \frac{\ag}{0.01} \right) \left( \frac{0.1}{\hout} \right)^2
\nonumber \\
&\times \frac{\bMb^2 \brout^{9/2}}{\bab^6 \bMst^{3/2}} \left( \frac{2\pi}{\text{yr}} \right),
\label{eq:cgb} \\
\cgs = \, &2.04 \times 10^{-10} \Gamma_{\rm s}  \left( \frac{\ag}{0.01} \right) \left( \frac{0.1}{\hin} \right)^2 \left( \frac{1358 \, \brout}{ \brin} \right)^{p-1}
\nonumber  \\
&\times \left( \frac{\kq}{0.1} \right)^2 \frac{\bMst^{1/2} \bRst^4}{\brin^4 \brout^{3/2}} \left( \frac{\bOmst}{0.1} \right)^4 \left( \frac{2\pi}{\text{yr}} \right), 
\label{eq:cgs} \\
\cgbs = & -2.04 \times 10^{-10} \Gamma_{({\rm bs})}  \left( \frac{\ag}{0.01} \right) \left( \frac{0.1}{\hout} \right)^2 \left( \frac{1358 \,  \brout}{\brin} \right)^{p-1}
\nonumber \\
&\times \left( \frac{\kq}{0.1}\right) \frac{\bMb \bRst^2 \brout^{1/2}}{\bMst^{1/2} \bab^3 \brin^2} \left( \frac{\bOmst}{0.1} \right)^2 \left( \frac{2\pi}{\text{yr}} \right),
\label{eq:cgbs}
\end{align}
where $\hin = (\rin/\rout)^{q-1/2} \hout$.  The rescaling above has removed the strongest dependencies of the damping rates on $p$, $q$, and $\rin/\rout$.  Table~\ref{tab:visc} lists values of the dimensionless viscous coefficients $\Gamma_{\rm b}$, $\Gamma_{\rm s}$, and $\Gamma_{({\rm bs})}$, varying $p$ and $q$.

Note that there are ``mixed'' terms in Eqs.~\eqref{eq:dldt_visc}-\eqref{eq:dsdt_visc}: the counter-aligment rate of $\ld$ and $\blb$ depends on $\bs$, while the counter-alignment rate of $\ld$ and $\bs$ depends on $\blb$.  Also note that net spin-disk alignment rate is given by
\be
\gamma_{\rm sd} = \left(1 + \frac{\Ld}{S} \right) \gamma_{\rm s}.
\label{eq:cgsd}
\ee
Assuming $\Ld \gg S$, $\gamma_{\rm sd}$ evaluates to be
\begin{align}
\gamma_{\rm sd} \simeq \ &7.52 \times 10^{-9} \frac{(2-p)\Gamma_{\rm s}}{5/2-p} \left( \frac{\ag}{0.01} \right) \left( \frac{0.1}{h_{\rm in}} \right)^2 \left( \frac{1358 \brout}{\brin} \right)^{p-1}
\nonumber \\
&\times \left( \frac{2 \kq}{\ks} \right) \left( \frac{\kq}{0.1} \right) \frac{\bMd \bRst^{7/2}}{\bMst^{1/2} \brin^4 \brout} \left( \frac{\bOmst}{0.1} \right)^3 \left( \frac{2\pi}{\text{yr}} \right).
\end{align}

Figure~\ref{fig:cgb} plots the disk-binary damping rate $\cgb$ as a function of the binary semi-major axis $\ab$.  In agreement with \cite{FoucartLai(2014)}, we find the damping rate to be small, and weakly dependent on the power-law surface density and sound-speed indices $p$ and $q$.  This is because the torque from the binary companion is strongest around $r \sim \rout$.  The properties of the disk near $\rout$ are ``global,'' since the amount of inertia of disk annuli near $\rout$ is set mainly by the total disk mass rather than the surface density profile, and the disk sound-speed does not vary greatly around $r \sim \rout$.  We conclude that viscous torques from disk warping are unlikely to significantly decrease the mutual disk-binary inclination $\tgdb$ unless $\ab \lesssim 200 \, \text{au}$.

Figure~\ref{fig:cgsd} plots the star-disk alignment rate $\cgsd$ as a function of the normalized stellar rotation frequency $\bOmst$.  Unlike the disk-binary alignment rate $\cgb$ (Fig.~\ref{fig:cgb}), $\cgsd$ depends strongly on the surface density and sound-speed power-law indices $p$ and $q$.  The alignment rate of a circumbinary disk with its binary orbital plane has a similarly strong dependence on $p$ and $q$ \citep{FoucartLai(2013),FoucartLai(2014),LubowMartin(2018)}.  This strong dependence arises because the torque on the inner part of a disk from an oblate star or binary is strongest near $\rin$.  The disk properties near $r \sim \rin$ are very local (both the amount of inertia for disk annuli and disk sound-speed), and hence will depend heavily on $p$ and $q$.  Despite this uncertainty, Figure~\ref{fig:cgsd} shows that there are reasonable parameters for which viscous torques from disk warping can significantly reduce the star-disk inclination $\tgsd$ [when $\cg_{\rm sd} \gtrsim 0.1 (2\pi/\text{Myr})$], especially when the stellar rotation rate is sufficiently high ($\bOmst \gtrsim 0.2$).

\section{Evolution of the Star-Disk-Binary System with Viscous Dissipation from Disk Warping}
\label{sec:Dyn}

\begin{figure*}
\centering
\includegraphics[scale=0.5]{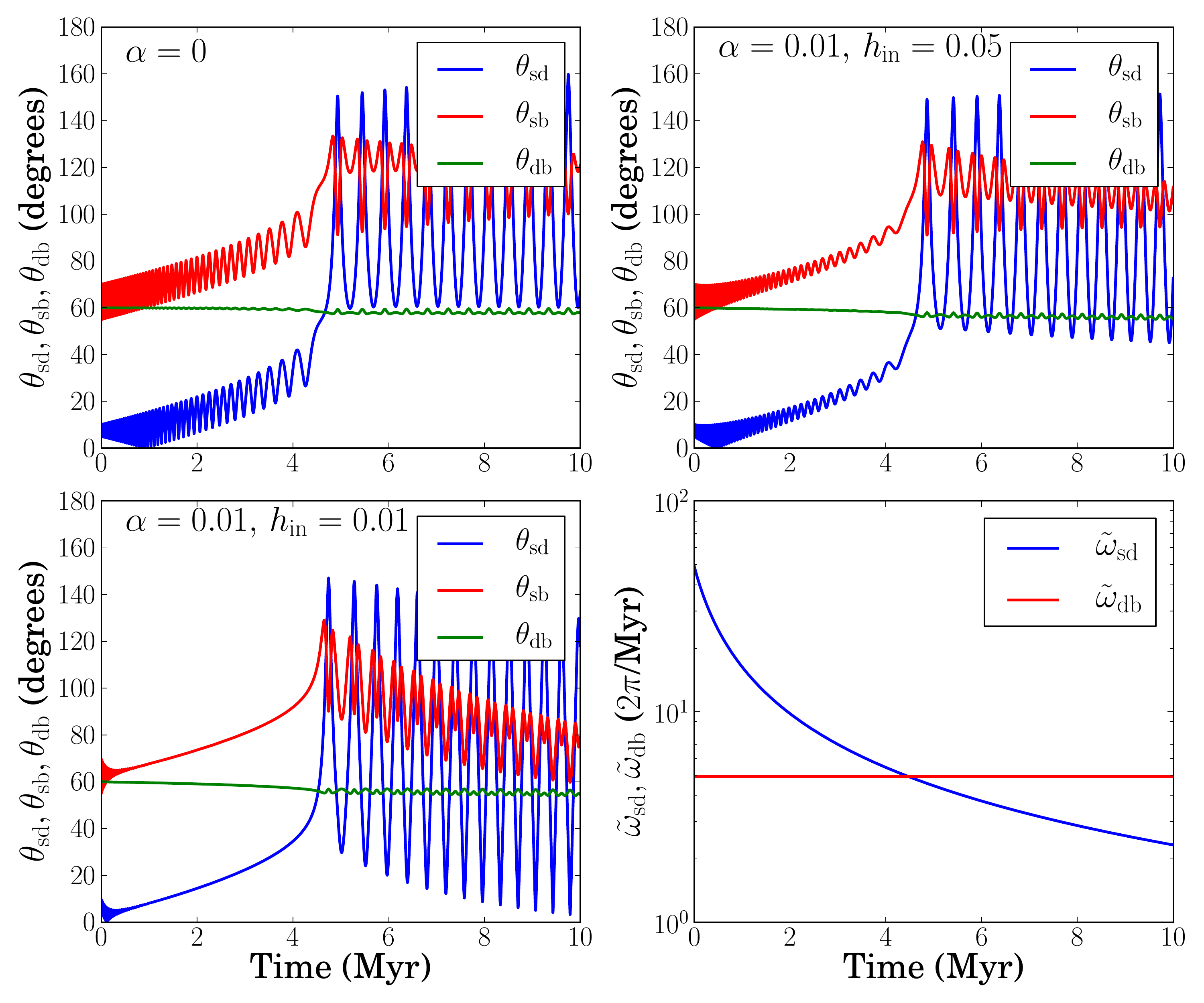}
\caption{
Inclination evolution of star-disk-binary systems.  The top panels and bottom left panel plot the time evolution of the angles $\tgsd$ [Eq.~\eqref{eq:tgsd}], $\tgsb$ [Eq.~\eqref{eq:tgsb}], and $\tgdb$ [Eq.~\eqref{eq:tgdb}], integrated using Eqs.~\eqref{eq:dsdt} and~\eqref{eq:dldt}, with values of $\ag$ and $h_{\rm in}$ [Eq.~\eqref{eq:q}] as indicated.  The bottom right panel shows the precession frequencies $\bomsd$ [Eq.~\eqref{eq:bomsd}] and $\bomdb$ [Eq.~\eqref{eq:bomdb}].  We take $\tgdb(0) = 60^\circ$, $\tgsd(0) = 5^\circ$, and $h_{\rm out} = 0.05$ [Eq.~\eqref{eq:q}].  The damping rates are $\gamma_{\rm b} = 5.05 \times 10^{-9} (2\pi/\text{yr})$ [Eq.~\eqref{eq:cgb}], $\gamma_{\rm sd}(0) = 2.00 \times 10^{-7} (2\pi/\text{yr})$ [Eq.~\eqref{eq:cgsd}], and $\gamma_{\rm bs} = -8.18 \times 10^{-10} (2\pi/\text{yr})$ [Eq.~\eqref{eq:cgbs}] for $h_{\rm in} = 0.05$, and $\gamma_{\rm b} = 7.12 \times 10^{-9} (2\pi/\text{yr})$, $\gamma_{\rm sd}(0) = 1.37 \times 10^{-6} (2\pi/\text{yr})$, and $\gamma_{({\rm bs})} = -1.51 \times 10^{-9}(2\pi / \text{yr})$ for $h_{\rm in} = 0.01$.
}
\label{fig:tg_warp_far}
\end{figure*}

\begin{figure}
\centering
\includegraphics[scale=0.5]{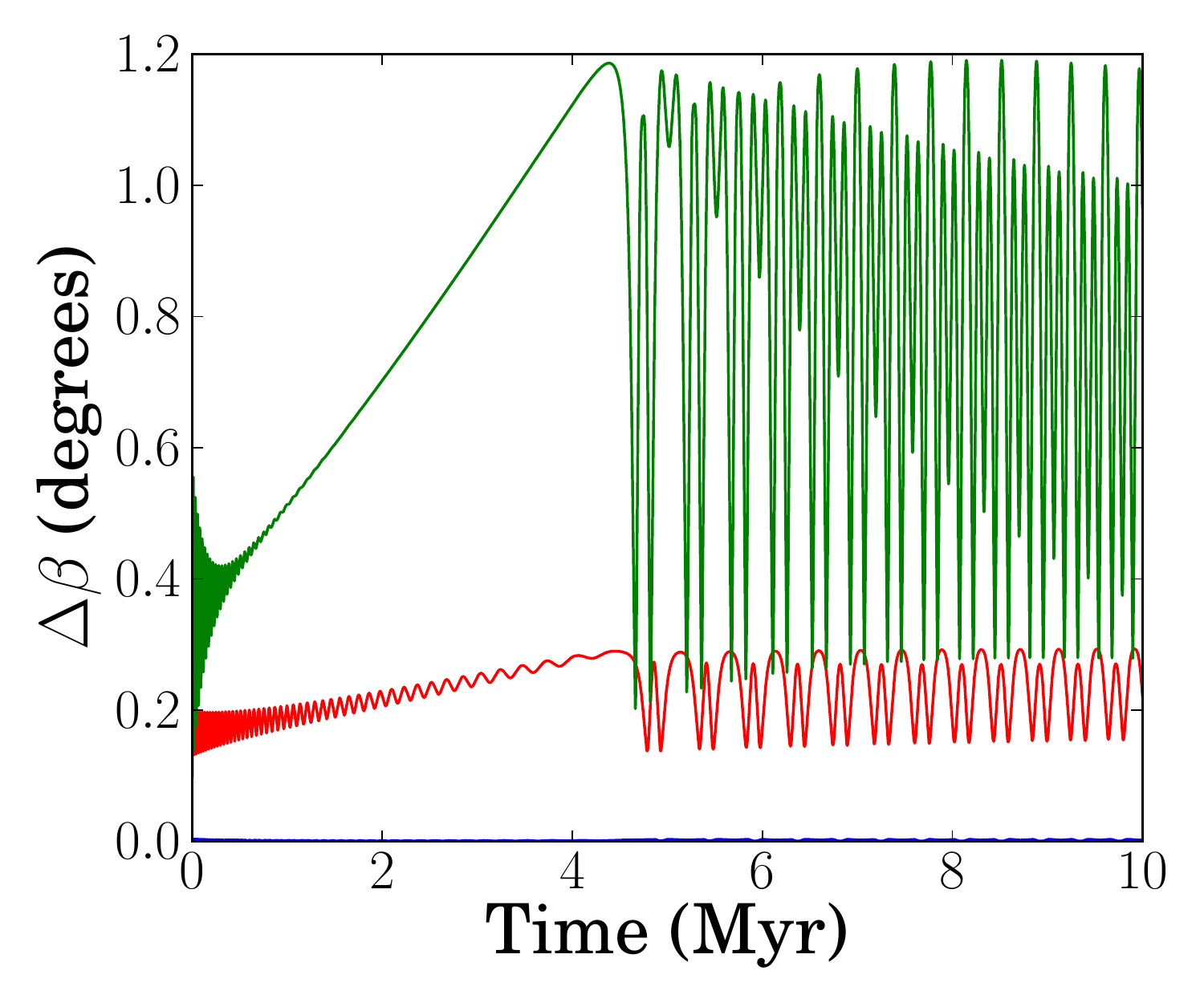}
\caption{ 
Total disk warp $\Delta \bg$ [Eq.~\eqref{eq:Dgbg}] for the integrations of Fig.~\ref{fig:tg_warp_far}.  The blue curve denotes the integration where $(\hin,\ag) = (0.05,0.0)$, the red is $(\hin,\ag) = (0.05,0.01)$, and the green is $(\hin,\ag) = (0.01,0.01)$.  All other parameters are listed in Fig.~\protect\ref{fig:tg_warp_far}.  All examples considered have $\Dg \bg < 1.2^\circ$, indicating the disk remains highly coplanar throughout the system's evolution.  Notice $\Dg \bg \ll 1^\circ$ when $\ag = 0$ (blue, hugs the x-axis).
}
\label{fig:warp_far}
\end{figure}

\begin{figure*}
\centering
\includegraphics[scale=0.5]{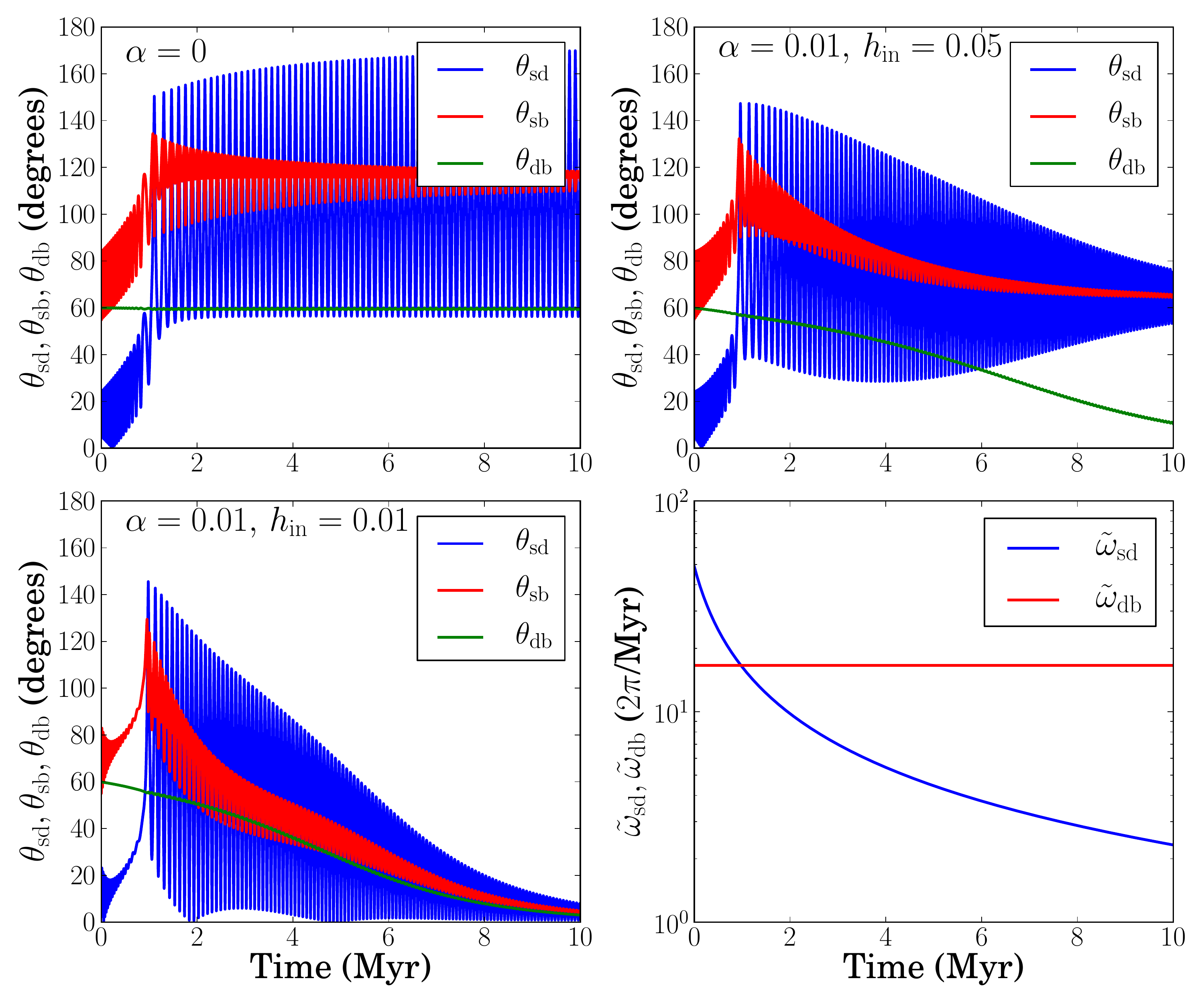}
\caption{
Same as Figure \ref{fig:tg_warp_far}, except $\ab = 200 \, \text{AU}$.  The damping rates are $\gamma_{\rm b} = 5.75 \times 10^{-8} (2\pi/\text{yr})$, $\gamma_{\rm sd}(0) = 2.00 \times 10^{-7} (2\pi/\text{yr})$, and $\gamma_{({\rm bs})} = -2.76 \times 10^{-9} (2\pi/\text{yr})$ for $h_{\rm in} = 0.05$, and $\gamma_{\rm b} = 8.11 \times 10^{-8} (2\pi/\text{yr})$, $\gamma_{\rm sd}(0) = 1.37 \times 10^{-6} (2\pi/\text{yr})$, and $\gamma_{({\rm bs})} = -5.10 \times 10^{-9}(2\pi / \text{yr})$ for $h_{\rm in} = 0.01$.
}
\label{fig:tg_warp_close}
\end{figure*}

\begin{figure}
\centering
\includegraphics[scale=0.55]{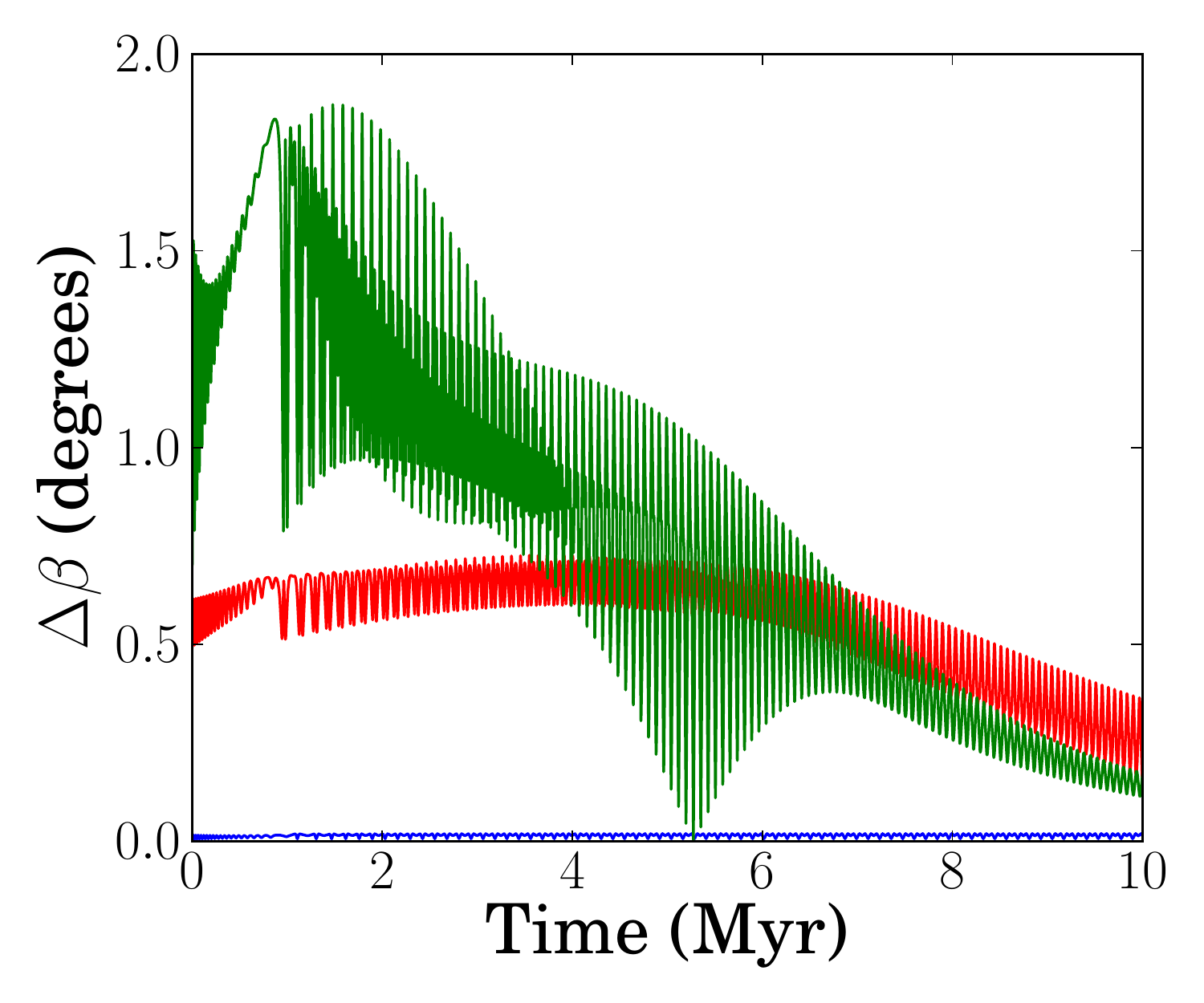}
\caption{
Same as Fig.~\ref{fig:warp_far}, except for the examples considered in Fig.~\ref{fig:tg_warp_close}.  All examples considered have $\Dg \bg <1.9^\circ$, indicating the disk remains highly coplanar throughout the disk's lifetime.
}
\label{fig:warp_close}
\end{figure}

This section investigates the evolution of star-disk-binary systems under gravitational and viscous torques:
\begin{align}
\frac{\der \bs}{\der t} = \ & -\bomsd (\bs \bcdot \ld)\ld \btimes \bs + \left( \frac{\der \bs}{\der t} \right)_{\rm visc},
\label{eq:dsdt} \\
\frac{\der \ld}{\der t} = \ & -\bomds (\ld \bcdot \bs) \bs \btimes \ld 
\nonumber \\
&- \bomdb (\ld \bcdot \blb) \blb \btimes \ld + \bigg( \frac{\der \ld}{\der t} \bigg)_{\rm visc}.
\label{eq:dldt}
\end{align}
The viscous terms are given by Eqs.~\eqref{eq:dldt_visc}-\eqref{eq:dsdt_visc}.  As in \cite{BatyginAdams(2013)} and \cite{Lai(2014)}, we assume the disk's mass is depleted according to
\be
M_{\rm d}(t) = \frac{M_{\rm d0}}{1+t/t_{\rm v}},
\label{eq:Md_time}
\ee
where $M_{\rm d0} = 0.1 \, \text{M}_\odot$ and $t_{\rm v} = 0.5 \, \text{Myr}$.  See \cite{Lai(2014)} and \cite{ZanazziLai(2017b)} for discussions on the dynamical evolution of $\bs$ and $\ld$ and secular resonance ($\bomsd \sim \bomdb$) when viscous dissipation from disk warping is neglected.

The effect of the $\cgs$ term on the dynamical evolution of $\bs$ over viscous timescales depends on the precessional dynamics of the star-disk-binary system.  If $\bomd \gg \bomb$, $\bs$ rapidly precesses around $\ld$, and the $\cgs$ term works to align $\bs$ with $\ld$.  If $\bomd \ll \bomb$, $\bs$ cannot ``follow" the rapidly varying $\ld$, and effectively precesses around $\blb$.  In the latter case, because of the rapid variation of $\ld$ around $\blb$, $\bs$ is only effected by the secular $\ld$.  As a result, $\cgs$ works to drive $\tgsb$ to $\tgdb$.  The effect of the $\cgb$ term is simpler: it always works to align $\ld$ with $\blb$.

Figure~\ref{fig:tg_warp_far} shows several examples of the evolution of star-disk-binary systems.  The top panels and bottom left panel of Fig.~\ref{fig:tg_warp_far} show the time evolution of the angles
\begin{align}
\tgsd &= \cos^{-1}(\bs \bcdot \ld),
\label{eq:tgsd} \\
\tgsb &= \cos^{-1}(\bs \bcdot \blb),
\label{eq:tgsb} \\
\tgdb &= \cos^{-1}(\ld \bcdot \blb),
\label{eq:tgdb}
\end{align}
from integrating Eqs.~\eqref{eq:dsdt}-\eqref{eq:dldt}, while the bottom right panel plots the characteristic precession frequencies $\bomd$ and $\bomb$.  The top left panel of Fig.~\ref{fig:tg_warp_far} does not include viscous torques ($\ag = 0$).  Because the damping rates $\cgb$ [Eq.~\eqref{eq:cgb}] and $\cgsd$ [Eq.~\eqref{eq:cgsd}] are much less than $0.1(2\pi/\text{Myr})$ over most of the system's lifetime (10 Myr), viscous torques have a negligible effect on the evolution of $\tgsd$, $\tgsb$, and $\tgdb$.  The bottom left panel of Fig.~\ref{fig:tg_warp_far} shows the evolution of $\tgsd$, $\tgsb$, and $\tgdb$ with $\ag = 0.01$ and $\hin = 0.01$.  Because the inner edge of the disc has a much smaller scaleheight, the oblate star warps the inner edge of the disk more [Eq.~\eqref{eq:est_twist}], resulting in $\cgsd$ taking a value larger than $0.1(2\pi/\text{Myr})$.  This increase in $\cgsd$ causes a much tighter coupling of $\bs$ to $\ld$ before secular resonance ($\bomd \gtrsim \bomb$), evidenced by the damped oscillations in $\tgsd$.  After secular resonance ($\bomd \lesssim \bomb$), the $\cgs$ term damps $\bs$ toward $\blb$.  Notice $\tgsb$ approaches $\tgdb$ because of the rapid precession of $\ld$ around $\blb$ after secular resonance, not $\tgsb \to 0$.

To gain insight to how the disk warp evolves during the star-disk-binary system's evolution, we introduce the misalignment angle $\Dg \bg$ between the disk's outer and inner orbital angular momentum unit vectors:
\begin{align}
&\sin \Dg \bg(t) = \big| \bl(\rout,t) \btimes \bl(\rin,t) \big|
\nonumber \\
&\simeq \big| [\blone(\rout,t) - \blone(\rin,t)]\btimes \ld(t) \big|
\label{eq:Dgbg}
\end{align}
Figure~\ref{fig:warp_far} plots $\Dg \bg$ as a function of time, for the examples considered in Fig.~\ref{fig:tg_warp_far}.  We see even when viscous torques from disk warping significantly alter the star-disk-binary system dynamics (e.g. $\ag = 0.01$ and $h_{\rm in} = 0.01$), $\Dg \bg < 1.2^\circ$ over the disk's lifetime, indicating a high degree of disk coplanarity throughout the system's evolution.

Figure~\ref{fig:tg_warp_close} is identical to Fig.~\ref{fig:tg_warp_far}, except we take $\ab = 200 \, \text{au}$ instead of $\ab = 300 \, \text{au}$.  Since $\cgb$ is greater than $0.1(2\pi/\text{Myr})$, $\ld$ aligns with $\cgb$ over the disk's lifetime.  In the top right panel of Fig.~\ref{fig:tg_warp_close}, $\cgsd$ is less than $0.1 (2\pi/\text{Myr})$ for most of the disk's lifetime, so $\bs$ stays misaligned with both $\ld$ and $\blb$.  At the end of the disk's lifetime, $\bs$ precesses around $\blb$, which is aligned with $\ld$.  In the bottom left panel, both $\cgb$ and $\cgsd$ are greater than $0.1(2\pi/\text{Myr})$ for most of the disk's lifetime.  This results in alignment of $\ld$, $\bs$, and $\blb$ over 10 Myr.  Figure~\ref{fig:warp_close} shows the evolution of disk misalignment angles for the examples considered in Fig.~\ref{fig:tg_warp_close}.  We see $\Dg \bg < 1.9^\circ$ for all examples considered, indicating the disk remains highly co-planar throughout the system's evolution.

\section{Discussion}
\label{sec:Discuss}

\subsection{Theoretical Uncertainties}

Our study of warped disks in star-disk-binary systems relies critically on the warp evolution equations derived in \cite{LubowOgilvie(2000)} for disks in the bending wave regime ($\ag \lesssim H/r$), assuming a small disk warp ($| \pd \bl/\pd \ln r| \ll 1$).  A non-linear disk warp will change the surface density evolution  of the disk through advection and viscosity where the warp is strongest (e.g. \citealt{Ogilvie(1999),TremaineDavis(2014)}).  In addition, even a small warp may interact resonantly with inertial waves, resulting in a parametric instability which enhances the disk's dissipation rate \citep{Gammie(2000),OgilvieLatter(2013)}.  Because we have found for typical parameters, the warp in the disk torqued externally by a central oblate star and distant binary is small [see Eqs.~\eqref{eq:tb_scale}-\eqref{eq:Wbb_scale}, \eqref{eq:ts_scale}-\eqref{eq:Wss_scale}, and~\eqref{eq:Wbs}-\eqref{eq:Wsb}], such effects are unlikely to change the main results of this paper.

In this study, we have assumed that the circumstellar disk in a binary system is circular.  This may not be a valid assumption, as the disk may undergo eccentricity growth through resonant Lindblad torques \citep{Lubow(1991)} or the Lidov-Kozai effect \citep{Martin(2014),Fu(2015a), ZanazziLai(2017a),LubowOgilvie(2017)}.  Lindblad torques only cause eccentricity growth where the binary orbital frequency is commensurate with the disk orbital frequency, so they are unlikely to be relevant unless the outer edge of the disk is close to tidal truncation by the binary companion.  Lidov-Kozai oscillations are a much more likely culprit for causing eccentricity growth of circumstellar disks in binaries when $\tgdb \gtrsim 40^\circ$.  Lidov-Kozai oscillations may be suppressed by the disk's self-gravity when \citep{Fu(2015b)}
\be
\Md \gtrsim 0.04 \, \Mb  \left( \frac{3 \rout}{\ab} \right)^3,
\ee
and by the disk's pressure gradients when \citep{ZanazziLai(2017a),LubowOgilvie(2017)}
\be
\ab \gtrsim 4.2 \, \rout \left( \frac{\Mb}{\Mst} \right)^{1/3} \left( \frac{\hout}{0.1} \right)^{-2/3}.
\ee
For our canonical parameters [Eq.~\eqref{eq:pars}], the Lidov-Kozai effect is unlikely to be relevant unless $\ab \lesssim 4 \rout$.

\subsection{Observational Implications}

In our companion work \citep{ZanazziLai(2017b)}, we show that the formation of a short-period (orbital periods less than 10 days) massive planet in many instances significantly reduces or completely suppresses primordial misalignments generated by the gravitational torque from an inclined binary companion.  Primordial misalignments are still robustly generated in protostellar systems forming low-mass ($\sim 1 \, \text{M}_\oplus$) multiple planets, and systems with cold (orbital periods greater than one year) Jupiters.  On the other hand, observations suggest that most Kepler compact multi-planet systems have small stellar obliquities (e.g. \citealt{Albrecht(2013),Winn(2017)}).  A major goal of this work was to examine if viscous torques from disk warping may reduce or suppress the generation of primordial misalignments in star-disk-binary systems.  We find that for some parameters, the star-disk inclination damping rate can be significant (see Fig.~\ref{fig:cgsd}); in particular, the star-disk misalignment may be reduced when the disk is sufficiently cold with strong external torques (Figs.~\ref{fig:tg_warp_far} \&~\ref{fig:tg_warp_close}).

Observational evidence is mounting which suggests hot stars (effective temperatures $\gtrsim 6000^\circ \, \text{K}$) have higher obliquities than cold stars \citep{Winn(2010),Albrecht(2012),Mazeh(2015),LiWinn(2016)}.  Since all damping rates from viscous disk-warping torques in star-disk-binary systems are inversely proportional to the disk's sound-speed squared [see Eqs.~\eqref{eq:cgb}-\eqref{eq:cgsd}], a tempting explanation for this correlation is that hot stars have hot disks with low damping rates which remain misaligned, while cold stars have cold disks with high damping rates which have star-disk misalignments significantly reduced over the disk's lifetime.  However, we do not believe this is a likely explanation, since the protostellar disk's temperature should not vary strongly with the T-Tauri stellar mass.  If a disk is passively heated from irradiation by its young host star \citep{ChiangGoldreich(1997)}, low mass ($\lesssim 3 \, \Msun$) pre-main sequence stars have effective temperatures which are not strongly correlated with their masses \citep{Hayashi(1961)}.  If the disk is actively heated by turbulent viscosity \citep{Lynden-BellPringle(1974)}, the disk's accretion rate does not vary enough between different host star masses to create a difference in disk temperature \citep{Rafikov(2017)}.

Even in systems where viscous torques from disk warping alter the dynamics of the star-disk-binary system over the disk's lifetime (Figs.~\ref{fig:tg_warp_far} \&~\ref{fig:tg_warp_close}), we find the misalignment angle between the outer and inner disk orbital angular momentum unit vectors to not exceed a few degrees (Figs.~\ref{fig:warp_far} \&~\ref{fig:warp_close}).  Therefore, it is unlikely that the disk warp profile plays a role in setting the mutual inclinations of forming exoplanetary systems with inclined binary companions.

\section{Conclusions}
\label{sec:Conc}

We have studied how disk warps and the associated viscous dissipation affect the evolution of star-disk inclinations in binary systems.  Our calculation of the disk warp profile shows that when the circumstellar disk is torqued by both the exterior companion and the central oblate star, the deviation of the disk angular momentum unit vector from coplanarity is less than a few degrees for the entire parameter space considered (Figs.~\ref{fig:warp_far} \&~\ref{fig:warp_close}).  This indicates that disk warping in star-disk-binary systems does not alter exoplanetary architectures while the planets are forming in the disk.  We have derived analytical expressions for the viscous damping rates of relative inclinations (Sec.~\ref{sec:Visc}), and have examined how viscous dissipation affects the inclination evolution of star-disk-binary systems.  Because the star-disk [Eq.~\eqref{eq:cgsd}, Fig.~\ref{fig:cgsd}] and disk-binary [Eq.~\eqref{eq:cgb}, Fig.~\ref{fig:cgb}] alignment timescales are typically longer than the protoplanetary disk's lifetime ($\lesssim 10 \, \text{Myrs}$), viscous dissipation from disk warping does not significantly modify the long-term inclination evolution of most star-disk-binary systems (Fig.~\ref{fig:tg_warp_far}, top left panel).  However, in sufficiently cold disks (small $H/r$) with strong external torques from the oblate star or inclined binary companion, the star-disk-binary evolution may be altered by viscous dissipation from disk warping, reducing the star-disk misalignment generated by star-disk-binary interactions (Figs.~\ref{fig:tg_warp_far} \&~\ref{fig:tg_warp_close}).  In particular, we find when the stellar rotatation rate is sufficiently high (rotation periods $\lesssim 2 \, \text{days}$), the star-disk damping is particularly efficient (Fig.~\ref{fig:cgsd}).  This viscous damping may explain the observed spin-orbit alignment in some multiplanetary systems (e.g. \citealt{Albrecht(2013),Winn(2017)}) in the presence of inclined binary companions.

\section*{Acknowledgements}

We thank the referee, Christopher Spalding, for many comments which improved the presentation and clarity of this work.  JZ thanks Re'em Sari and Yoram Lithwick for helpful discussions.  This work has been supported in part by NASA grants
NNX14AG94G and NNX14AP31G, and NSF grant AST-
1715246.  JZ is supported by a NASA Earth and Space Sciences Fellowship in Astrophysics.

\end{document}